# On the interpretation of the kinetics of ligand-receptor binding

David Colquhoun[1] and James P Higham[2]

[1] Neuroscience, Physiology and Pharmacology, University College London, Gower Street, London, WC1E 6BT, UK

[2] Department of Pharmacology, University of Cambridge, Tennis Court Road, Cambridge, CB2 1PD, UK

Correspondence may be addressed to d.colquhoun@ucl.ac.uk or jph87@cam.ac.uk

## Abstract

When the rates of ligand binding are measured by methods such as surface plasmon resonance, it's common practice to use the observed rate constants for the onset and offset of binding to estimate an equilibrium constant for ligand binding. If this agrees with the equilibrium constant found as the $EC_{50}$ for binding at equilibrium, this is taken as validation of the measured rates. This is correct only when binding produces no conformation change in the receptor, and ligand binding follows a single exponential time course. Here, we investigate a simple 3-state model in which binding is followed by a conformation change in the receptor. Three special cases of this model in which the time course of onset and offset of ligand binding are close to being single exponentials are analysed. These cases are (1) when binding is much faster than the conformation change, (2) when the conformation change is much faster than binding, and (3) when the rates of ligand dissociation and receptor activation are both fast. It's concluded that the measured rates will often yield an estimate of the equilibrium constant for ligand binding that is close to the effective, or macroscopic, equilibrium constant – the $EC_{50}$ found by measuring binding at equilibrium – which depends on both of the underlying microscopic equilibrium constants describing ligand binding and the conformation change. The exception to this conclusion is the case when binding is much faster than the subsequent conformation change, though the estimate of the equilibrium constant for ligand binding still depends on both of the underlying microscopic equilibrium constants.

## 1. Introduction

How rapidly a ligand interacts with its receptor is a central question in pharmacology, yet the kinetics of ligand binding do not always have a straightforward interpretation. This review is an updated version of “How fast do drugs work?” (1981) [1], which examined the kinetics of drug action. Since that was written, there have been advances in both the measurement of ligand binding, and in the measurement and interpretation of the activity of single molecules. The problem of interpreting reaction kinetics has been dealt with in a general way by Manfred Eigen [2] [3], in the 1960s, in the context of fast chemical relaxation methods. In the 1970s and 1980s, this general theory was applied to receptors [4–6]. We reiterate the method of calculating the time course of the occupancy of each state, and combinations of states, of a system. For the most part, we will focus on the relationship between observed (macroscopic) rates of ligand binding and the underlying microscopic rate constants which provide information about the ligand-receptor interaction.

## 2. The simplest case: binding to identical, independent binding sites

In 1909, AV Hill [7] derived equations describing the binding of a ligand to identical independent binding sites. He derived not only the fraction of sites that are occupied (the ‘occupancy’) at equilibrium, but also the rate at which equilibrium is approached. This derivation was done in the context of the binding of curare to muscle type nicotinic receptors. Hill did this nine years before Langmuir (1918) [8] did much the same in the context of the binding of gases to metal surfaces (electric light bulb filaments).

The binding of a ligand, A, to a binding site, R (e.g., a receptor), can be written thus

$$\mathrm{R} \underset{k_{-1}}{\overset{k_{+1}}{\rightleftharpoons}} \mathrm{AR} \tag{1}$$

The constants on the arrows are named rate constants: $k_{+1}$ is the association rate constant and $k_{-1}$ is the dissociation rate constant. The approach to equilibrium for the reaction in Scheme (1) follows a simple exponential time course, with a macroscopic rate constant, $\lambda$, given by

$$\lambda = k_{+1}x_{\mathrm{A}} + k_{-1} = \frac{1}{\text{mean lifetime in R}} + \frac{1}{\text{mean lifetime in AR}} \tag{2}$$

where $x_{\mathrm{A}}$ is the free ligand concentration, assumed to be constant. The change in occupancy with time as equilibrium is approached is described as the *relaxation* towards equilibrium: the system ‘relaxes’ to a state with low energy.

The time constant, $\tau$, for the exponential relaxation is the reciprocal of the rate constant, thus

$$\tau = \frac{1}{\lambda} = \frac{1}{k_{+1}x_{\mathrm{A}} + k_{-1}} \tag{3}$$

The units of $k_{-1}$ are reciprocal seconds, $\mathrm{s}^{-1}$, the units of rate or frequency. The units of the association rate constant, $k_{+1}$, are $\mathrm{M}^{-1}\mathrm{s}^{-1}$ where M is concentration (molar). Association of ligand will be faster at high concentrations than at lower concentrations: the transition rate for the association reaction will be $k_{+1}x_{\mathrm{A}}$, which will have units of: $\mathrm{s}^{-1}$. If the free concentration of the ligand is constant during binding process then $k_{+1}x_{\mathrm{A}}$ will be constant; it is often called a pseudo-first order rate constant.

If the occupancy at zero time is denoted $p(0)$, then the time course of the approach of the occupancy towards its equilibrium value, $p(\infty)$, is described by a single exponential term, with the occupancy at time *t* given by

$$p(t) = p(\infty) + [p(0) - p(\infty)]\exp\left(-\frac{t}{\tau}\right) \tag{4}$$

At time zero, *t* = 0, the exponential term is $e^0 = 1$ so this reduces to $p(0) = p(0)$. After a long time, as $t \to \infty$, the exponential term tends to zero, so Eq. (4) reduces to the equilibrium form, $p(\infty) = p(\infty)$, where the equilibrium occupancy is given by

$$p(\infty) = \frac{x_{\mathrm{A}}}{x_{\mathrm{A}} + K_{\mathrm{A}}} \equiv \frac{c_{\mathrm{A}}}{1 + c_{\mathrm{A}}} \tag{5}$$

where $x_{\mathrm{A}}$ is the free concentration of ligand, $K_{\mathrm{A}}$ is the dissociation equilibrium constant, defined as the ratio of the two microscopic rate constants, and $c_{\mathrm{A}}$ is defined as the normalised concentration, the ratio of ligand concentration to its equilibrium constant

$$K_{\mathrm{A}} \equiv \frac{k_{-1}}{k_{+1}}$$

$$c_{\mathrm{A}} \equiv \frac{x_{\mathrm{A}}}{K_{\mathrm{A}}} \tag{6}$$

The equilibrium constant, $K_{\mathrm{A}}$, has units of concentration – it is the concentration at which half the receptors are occupied according to Eq. (5). The normalised concentration, $c_{\mathrm{A}}$, is, therefore, dimensionless and $c_{\mathrm{A}} = 1$ corresponds to 50% occupancy; see, for example, [7] [9] [10] [11].

It's important to notice the rate constant for the observed time course of the approach equilibrium, Eq. (2), is not the same as either of the underlying microscopic rate constants. The observed rate constant, Eq. (2), is called the macroscopic rate constant and its value depends on all of the microscopic rate constants. There is, for any reaction mechanism, however simple or complicated, no straightforward relationship between the observed macroscopic rate constants for equilibration and the underlying microscopic rate constants that define the reaction mechanism.

The microscopic constants provide information about the reaction mechanism, but they aren't directly observable. The challenge is to untangle the relationship between the observed macroscopic rates and the interpretable microscopic rates. This has been achieved in the case of single ion channel measurements [12] [13] [14], but ligand binding experiments are generally treated as simple binding reactions as in Scheme (1).

It is common practice to start an experiment at zero concentration so $p(0) = 0$, Eq. (4). Then the concentration is increased as quickly as possible to a concentration, $x_{\mathrm{A}}$, and the onset of binding, or of channel opening, is observed. If the time course of this approach to equilibrium is a simple exponential, its rate constant can be measured and it is usually interpreted as in Eq. (2). In the binding literature, the rate constant for the onset of binding, so measured, is generally referred to as $k_{\mathrm{on}}$. From Eq. (2), this is interpreted as

$$k_{\mathrm{on}} = k_{+1}x_{\mathrm{A}} + k_{-1} \tag{7}$$

The dissociation of the ligand is then measured by switching the free ligand concentration to zero and following the dissociation of bound ligand over time. Insofar as the time course of dissociation can be described by a simple exponential curve, its rate is measured as $k_{\mathrm{off}}$, and the offset rate is interpreted, since $x_{\mathrm{A}} = 0$ in Eq. (7), as

$$k_{\mathrm{off}} = k_{-1} \tag{8}$$

Substitution of $k_{-1}$ so found into Eq. (7) allows the association rate constant, $k_{+1}$, to be found. The ratio of these rate constants, according to Eq. (6), provides an estimate of the dissociation equilibrium constant. In this simple case, the observed values of $k_{\mathrm{on}}$ and $k_{\mathrm{off}}$ can be used to estimate an equilibrium constant using Eq. (9). The value calculated from Eq. (9) we define as $K_{\mathrm{app}}$ because, although in this simple case it gives the true equilibrium constant, $K_{\mathrm{A}}$, this will not always be true.

$$K_{\mathrm{app}} \equiv \frac{k_{\mathrm{off}}x_{\mathrm{A}}}{k_{\mathrm{on}} - k_{\mathrm{off}}} \tag{9}$$

If the application of ligand is long enough for equilibrium to be attained, then the measured value for equilibrium occupancy can also be used to estimate $K_{\mathrm{A}}$, that is, the concentration needed to achieve 50% occupancy at equilibrium, via Eq. (5). If these two estimates of $K_{\mathrm{A}}$ are much the same, this is taken as a validation of the analysis. Whether or not this is a valid procedure will be discussed below.

Notice that the concentration of free ligand is assumed to be constant throughout the time over which occupancy is changing. This will be true only insofar as the concentration is changed in a stepwise fashion from 0 to $x_{\mathrm{A}}$ for the onset measurement, and again in a stepwise fashion from $x_{\mathrm{A}}$ to 0 for the offset measurement. This is referred to as a concentration jump experiment. But it isn't possible to change the concentration instantaneously, so this is always an approximation. The approximation will be good if the concentration change is completed in a time that is short compared with the rate of change of occupancy. The time taken for change of concentration in surface plasmon resonance experiments is at best about 100 ms (Biacore 2000 Instrument Handbook), whereas concentration changes in outside-out patches can be achieved in about 100 μs [15].

The theory that has been described so far is valid for "silent" competitive antagonists, *i.e.*, molecules that don't produce a conformation change in their receptor. But all agonists (and many antagonists that are actually weak partial agonists or inverse agonists) do produce a conformation change in the receptor, so they can't be described by the simple mechanism in Scheme (1). In the next section, we will describe how to interpret the kinetics of binding for a ligand that produces a conformation change in its receptor which will make clear that such cases should not be treated as simple binding reactions.

## 3. The simplest case in which binding produces a conformation change

It is obvious that binding of an agonist must result in a conformation change in its receptor. How else could its binding result in a biological response? It was not until 1957 that del Castillo and Katz [16] added the conformation change explicitly, surprisingly late, given that their model is analogous with the Michaelis and Menten model for enzymes given in 1913 [17]. For some history, see [18]. The del Castillo-Katz (hereinafter referred to as dCK) mechanism can be written thus

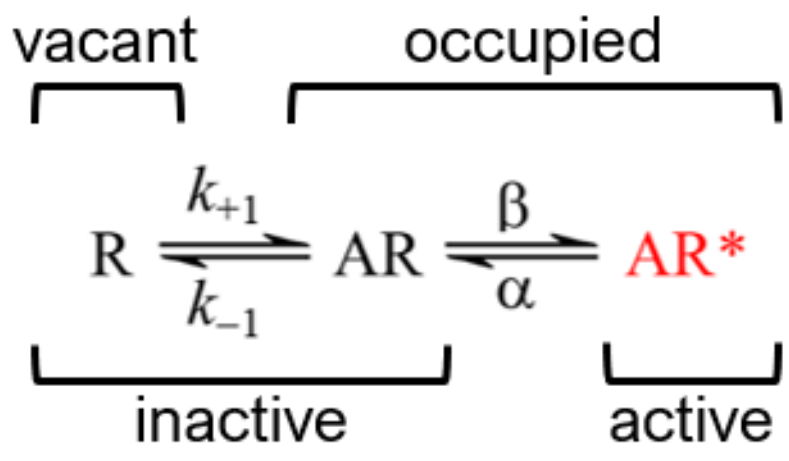

(10)

The ligand, A, binds to a receptor, R, to form a complex, AR, as in Eq. (1). But then the complex can change conformation (isomerise) to the active state, denoted AR*. In the particular case of an ion channel, inactive means shut and active means open.

The rate constants for the transitions between the three states of the receptor are marked on the arrows in Scheme (10). For the binding step, the equilibrium dissociation constant will be exactly as in Eq. (6)

$$K_A = \frac{k_{-1}}{k_{+1}} \tag{11}$$

with units of concentration. Its reciprocal, the equilibrium association constant, is often termed the affinity constant.

The equilibrium constant for the isomerisation step, the change in conformation of the ligand-receptor complex from the inactive to active conformation, can be defined as the ratio of the forward rate constant, $\beta$, to the rate constant for the reverse direction, $\alpha$. This equilibrium constant is often denoted *E*,

$$E = \frac{\beta}{\alpha} \tag{12}$$

The symbol *E* is used here because the value of this equilibrium constant clearly measures the property of agonists that Stephenson had described in 1956 as *efficacy*. Although del Castillo and Katz were unaware of Stephenson's work, their motive for suggesting that activation should be regarded as a separate step was similar to Stephenson's. The motive was to provide a mechanism that could account for the observation that some agonists could not produce a maximum response (the response at a saturating concentration) as large as the maximum response evoked by other agonists. These weaker agonists were referred to as *partial agonists.* At a very high concentration of agonist, all binding sites will be occupied and so the only states that exist are AR and AR*. Of these, the fraction that are active is $E/(1+E)$ so if, for example, $E=1$, then the maximum fraction of active receptors would be 0.5. The equilibrium properties of the dCK model are discussed in greater detail in, for example, [9] [10] [11]. Unfortunately, Stephenson failed to distinguish between microscopic and macroscopic equilibrium constants, so the methods that he proposed to measure affinity and efficacy don't work [19].

### 3.1. The rate of approach to equilibrium

The dCK mechanism is too simple to describe any real agonist-receptor interaction. It postulates the existence of three states of the receptor, described by four free rate constants. In contrast, the description of the activation of glycine-activated ion channels needs 10 discrete states and 14 free rate constants [20]. Nonetheless, the dCK mechanism is sufficient to illustrate the problems that occur when trying to infer mechanisms from observations of the kinetics of ligand binding.

Although the dCK mechanism is simple enough for the rates of approach to equilibrium to be worked out by hand, this is impractical for more complex mechanisms. The only way to do the calculations in general is to use the language of matrix algebra. Fortunately, what you need to know is not very daunting. If you don't know anything about matrices, you might find helpful to try this video tutorial, *Basic Matrix Algebra in 45 Minutes* [21], before tackling more detailed texts.

The beauty of matrix notation is illustrated by the fact that the rates of macroscopic equilibration for *any* mechanism can be described by the elegantly simple equation (Eq. (23) in ref. [5]),

$$\boldsymbol{p}(t) = \boldsymbol{p}(0)\exp(\boldsymbol{Q}t) \tag{13}$$

This simple equation is all that's needed for all the calculations in this paper. In Eq. (13), $\boldsymbol{p}(t)$ is a $1 \times k$ row matrix that contains the occupancies of each of the $k$ states at time $t$, and $\boldsymbol{Q}$ is the transition rate matrix (bold type indicates a matrix), a square $k \times k$ matrix, where $k$ is the number of discrete states of the receptor – for the dCK mechanism, $k = 3$. The states must be numbered, and, although the numbering is arbitrary, we'll follow the convention of Colquhoun and Hawkes, that numbering should start with the active states, so, for the dCK mechanism, state 1 is the active state, AR*, state 2 is the inactive but occupied state, AR, and state 3 is the vacant receptor, R. The solution in Eq. (13) holds only in the case where $\boldsymbol{Q}$ is constant, *e.g.*, the concentration of free ligand is constant.

Although Eq. (13) can be used directly in computer languages that have the ability to calculate the exponential of a matrix in a single command, this is an inefficient way to do the calculations [22]. It is more efficient, and clearer, to write Eq. (13) without matrices as the sum of $k - 1$ exponential components, each with its own time constant and amplitude. The numerical evaluation of it in this scalar form needs a few tricks. These are explained and exemplified elsewhere [23] [24].

For the dCK mechanism, with $k = 3$ states, the solution for the time course of the approach to equilibrium of the occupancy of the $j\text{th}$ state can be written, without using matrix notation, in the form of the sum of two exponential components. The macroscopic rate constants are denoted $\lambda_2$ and $\lambda_3$, and the amplitudes of these components at $t = 0$ are denoted $w_{2j}$ and $w_{3j}$ (the absence of $\lambda_1$ is explained below). So,

$$p_j(t) = p_j(\infty) + w_{2j} \exp(-\lambda_2 t) + w_{3j} \exp(-\lambda_3 t) \tag{14}$$

This gives the occupancy of a single state, $j$, over time. Eq. (13) gives the occupancies of each of the $k$ states at time $t$ in $\boldsymbol{p}(t)$, which is a $1 \times k$ matrix, a row vector, which is defined in the case of $k = 3$ as

$$\boldsymbol{p}(t) = [p_1(t) \quad p_2(t) \quad p_3(t)] \tag{15}$$

where $p_j(t)$ is the occupancy of the $j\text{th}$ state at time $t$. For example, with the numbering above, $p_1(t)$ is the time course of the fraction of receptors in the active state, AR*, and so on. The occupancies of each state at time zero are in $\boldsymbol{p}(0)$, and the equation describes the time course of the occupancy of each state from their values at time zero to their eventually-attained equilibrium values, $\boldsymbol{p}(\infty)$.

The transition rate matrix, $\boldsymbol{Q}$, is constructed thus. The element, $q_{ij}$, in the $i\text{th}$ row and the $j\text{th}$ column of $\boldsymbol{Q}$ is the transition rate (units of $\text{s}^{-1}$) from state $i$ to state $j$. The diagonal elements, those for which $i = j$, are constructed so that the sum of each row is zero.

### 3.1.1. The simple binding reaction revisited

Before considering the time course of agonist binding for the dCK mechanism in more detail, we will first briefly revisit the simple binding reaction, depicted in Scheme (1), which has $k = 2$ states. If we number the occupied state, AR, as state 1, and the vacant state, R, as state 2, then we have

$$\boldsymbol{Q} = \begin{bmatrix} -k_{-1} & k_{-1} \\ k_{+1}x_{\mathrm{A}} & -k_{+1}x_{\mathrm{A}} \end{bmatrix} \tag{16}$$

For example, the transition rate, $k_{-1}$, from state 1 (occupied) to state 2 (vacant), *i.e.*, the dissociation reaction, is $q_{12}$, the element in the first row and second column.

The time course of binding for Scheme (1) has the form given in Eq. (4); a single exponential time course with a time constant, $\tau$, and a rate constant, $\lambda = 1/\tau$.

The observed, or macroscopic rate constants can be found as eigenvalues, values that can be calculated from any square matrix; a $k \times k$ matrix has $k$ eigenvalues. Every computer language has the ability to calculate them. In simple cases, like $k = 2$ or $k = 3$, it's possible to write expressions to calculate their values but, in general, this isn't possible, though they can always be computed numerically. The eigenvalues of $-\boldsymbol{Q}$ provide the macroscopic rate constants for the approach of binding to equilibrium. The eigenvalues of $\boldsymbol{Q}$ are negative (or zero); the minus sign makes the eigenvalues positive. The two eigenvalues of $-\boldsymbol{Q}$ for the simple binding reaction, Eq. (16), are

$$\lambda_1 = 0 \tag{17}$$

$$\lambda_2 = k_{+1}x_{\mathrm{A}} + k_{-1} \tag{18}$$

Here, $\lambda_2$ provides the macroscopic rate constant, as in Eq. (2). The other eigenvalue is zero. This corresponds to the equilibrium occupancy when $t \to \infty$ in Eq. (4), as shown in Eq. (5). Since $e^0 = 1$, the equilibrium term, $p(\infty)$, can be written as $p(\infty)\exp(-\lambda_1 t)$. In other words, the equilibrium occupancy doesn't change with time (zero rate, or infinite time constant). This is explained further elsewhere [23] [24]. As a result of the way in which the $\boldsymbol{Q}$ matrix is constructed, with rows that sum to zero, it is singular and, hence, will always have one eigenvalue equal to zero.

So, the simplest case of ligand binding exemplifies that the macroscopic rate constants for the equilibration of binding are given by the non-zero eigenvalues of the $-\boldsymbol{Q}$ matrix. This can be applied to any receptor mechanism.

### 3.1.2. The del Castillo-Katz mechanism

If we number the states in Scheme (10) thus

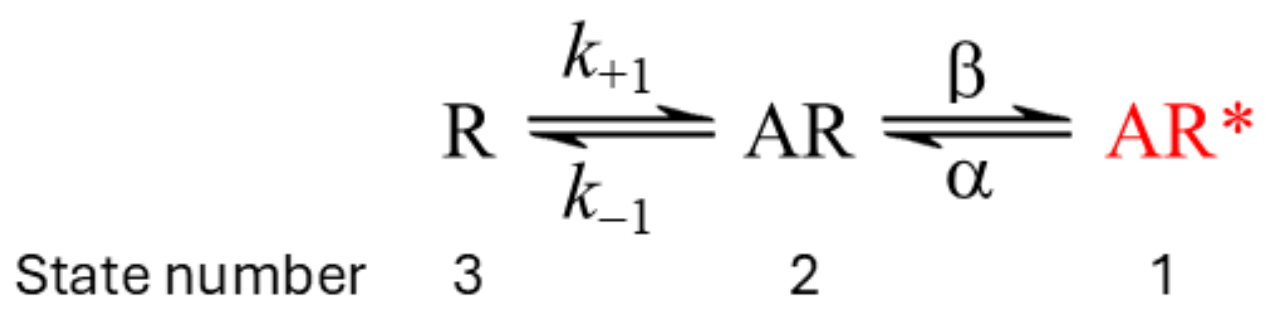

(19)

then the $\boldsymbol{Q}$ matrix for the dCK mechanism can be written as

$$\boldsymbol{Q} = \begin{bmatrix} -\alpha & \alpha & 0 \\ \beta & -(\beta + k_{-1}) & k_{-1} \\ 0 & k_{+1}x_{\mathrm{A}} & -k_{+1}x_{\mathrm{A}} \end{bmatrix} \tag{20}$$

The observed macroscopic rate constants for re-equilibration are the eigenvalues of $-\boldsymbol{Q}$. One of the eigenvalues, say $\lambda_1$ (the numbering of the eigenvalues is arbitrary), of Eq. (20) is zero, as usual, so

$$\lambda_1 = 0 \tag{21}$$

The other two, non-zero, eigenvalues, $\lambda_2$ and $\lambda_3$, can be found as the two roots of a quadratic equation

$$\lambda^2 - b\lambda + c = 0 \tag{22}$$

where

$$b = \lambda_2 + \lambda_3 = \alpha + \beta + k_{-1} + k_{+1}x_{\mathrm{A}} \tag{23}$$

$$c = \lambda_2\lambda_3 = \alpha k_{-1} + \alpha k_{+1}x_{\mathrm{A}} + \beta k_{+1}x_{\mathrm{A}} \tag{24}$$

Solving the quadratic gives

$$\lambda_2, \lambda_3 = \frac{1}{2}\left(b \pm \sqrt{b^2 - 4c}\right) \tag{25}$$

Again, we see that the microscopic rate constants (the ones in which we're interested) are related only indirectly to the observed (macroscopic) rate constants for equilibration. This makes it harder to estimate interpretable physical constants from observations of ligand binding. It's true for any square matrix that the sum of all the eigenvalues (the macroscopic rate constants) is equal to the sum of its diagonal elements and this, for the $\boldsymbol{Q}$ matrix, is the sum of all the microscopic rate constants.

To complete the solution, as given in Eq. (14), we need to calculate the amplitudes of the two components at $t = 0$, $w$. This can be done, for any mechanism, using the eigenvectors of the $\boldsymbol{Q}$ matrix. As well as having $k$ eigenvalues, the $\boldsymbol{Q}$ matrix also has $k$ eigenvectors – one for each eigenvalue. Every computer language has the ability to calculate the eigenvectors, but the mathematical details are not important for our purposes. Finding the amplitudes of the exponential components involves a beautiful trick known as spectral expansion, which provides the spectral matrices, $\boldsymbol{A}_{\mathrm{m}}$, of $\boldsymbol{Q}$, the elements of which are used to calculate the amplitudes of the exponential components. The spectral matrices, of which there are $k$, are derived from the eigenvectors of $\boldsymbol{Q}$. The details of how to calculate the amplitudes are described elsewhere [23] [24] [25], so they won't be repeated here. The method described in these references is implemented in the R program, `CK-kinetics.R` (see the electronic supplementary material), used to calculate the examples below.

The solution in Eq. (14) allows the calculation of the time course of the occupancies of the three states of the receptor. The active state, AR*, is defined by the numbering in Scheme (19) as state 1, so the time course of equilibration of the active state, $p_1(t)$, is given by Eq. (14) with $j = 1$, that is

$$p_1(t) = p_1(\infty) + w_{21}\exp(-\lambda_2 t) + w_{31}\exp(-\lambda_3 t) \tag{26}$$

Here, $\lambda_2$ and $\lambda_3$ are given in Eq. (23), (24) and (25). Similarly, the time course for the occupancy of the intermediate state, AR, is given by Eq. (14) with $j = 2$, and the time course for the occupancy of the resting (vacant) state, R, is given by Eq. (14) with $j = 3$. These are all given in the printout file of the R program (see the electronic supplementary material).

Ligand is bound to both states 1, AR*, and to state 2, AR, so the time course of ligand binding, $p_{\text{bound}}(t)$, is the sum of the occupancies of states 1 and 2, thus

$$p_{\text{bound}}(t) = p_1(t) + p_2(t) \tag{27}$$

This result can be generalised to cater for mechanisms in which more than one ligand can be bound to each receptor (as is the case for ligand-gated ion channels). Define $n_j$ as the number of ligand molecules bound to state $j$ of the receptor and $N$ as the total number of binding sites. For example the glycine receptor has $N = 3$ because 3 molecules of glycine must be bound to open the channel effectively [20]. The general result can be written

$$p_{\text{bound}}(t) = \frac{1}{N}\sum_{j=1}^{k} n_j p_j(t)$$

$$j = 1,2 \cdots k$$

$$n_j = 0,1, \cdots N \tag{28}$$

The maximum value of $j$ is the number of states in the mechanism ($k$), and the maximum value of $n_j$ is the number of binding sites per receptor ($N$). The division by $N$ in Eq. (28) is necessary to ensure that $p_{\text{bound}}(t) = 1$ at very high ligand concentration.

This result is analogous to the way in which the macroscopic current flowing through a population of ion channels is found by summing the occupancies of the open states, weighted by the conductance of each state (see [23], Section 4.6).

As usual, the result can be written more briefly in matrix notation. Define a $k \times 1$ column vector that contains the values of $n_j$. For the dCK mechanism, with $k = 3$, this would be

$$\boldsymbol{n} = \begin{bmatrix} 1 \\ 1 \\ 0 \end{bmatrix} \tag{29}$$

*i.e.*, one ligand molecule is bound to states 1 and 2 and none are bound to state 3. With this definition, we can write Eq. (28) as

$$p_{\text{bound}}(t) = \frac{1}{N}\boldsymbol{p}(t)\boldsymbol{n} \tag{30}$$

where $\boldsymbol{p}(t)$ is a $(1 \times k)$ row vector that contains the occupancies for each state. Note that $\boldsymbol{p}(t)\boldsymbol{n}$ is scalar – the shapes of the vectors are $(1 \times k)(k \times 1) = (1 \times 1)$. This is the form used in the R script that accompanies this paper (see the electronic supplementary material).

Notice that the time constants are the same for every state. All that varies is the amplitude of each component. And this is true also for any linear combination of states, such as those in Eq. (27), (28) and (30). All of these have the same macroscopic rate constants, but different amplitudes, for each of the exponential components. This behaviour is characteristic of Markov processes.

The equilibrium binding curve for the dCK mechanism has exactly the same form as for the simple mechanism shown in Scheme (1), but the ligand concentration required to occupy half of the receptors is not $K_\mathrm{A}$, but rather it is

$$K_\mathrm{eff} = \frac{K_\mathrm{A}}{1 + E} \tag{31}$$

This effective, or macroscopic, equilibrium constant depends on both the microscopic affinity, $K_\mathrm{A}$, and on the efficacy, $E$. This is why it's impossible to estimate separately affinity and efficacy using equilibrium measurements, at least when efficacy is high [19] [9].

### 3.2. Special cases of the del Castillo-Katz mechanism

Ligand binding following a jump in ligand concentration is usually fitted with a single exponential curve for both onset and offset. For the binding of an agonist, two exponential components are predicted, even for the simplest possible mechanism, the dCK mechanism in Scheme (10). An example of the time course of the occupancies of all 10 states of a realistic mechanism for the glycine receptor is given in [26] (see Figure 6 therein and see also the accompanying correction [27]). In that example, each curve is calculated as the sum of 9 exponential components. This example can be replicated by the R program used to calculate the examples below, but it won't be discussed in this paper.

In practice, it's quite common to observe activation, *e.g.* $p_\mathrm{open}(t)$, curves and ligand binding curves that follow a single exponential time course, at least roughly. Therefore, we'll consider the three special cases of the dCK mechanism in which only one of the two exponential components can be detected. These cases can be named as the '*fast binding*', '*fast conformation change*' and '*low intermediate concentration*' cases. These cases are also discussed elsewhere ([25], see Section 3.5). There are two reasons why the time course of equilibration might appear to have only one component: one of the components might not be detectable because it has a small amplitude, or one component might not be detectable because it's too fast to be measured. Both of these phenomena will be illustrated below.

These three cases are illustrated at the single molecule level in Figure 1 (a version of which was used for many years for teaching in the Plymouth Microelectrode Techniques course; see [28]).

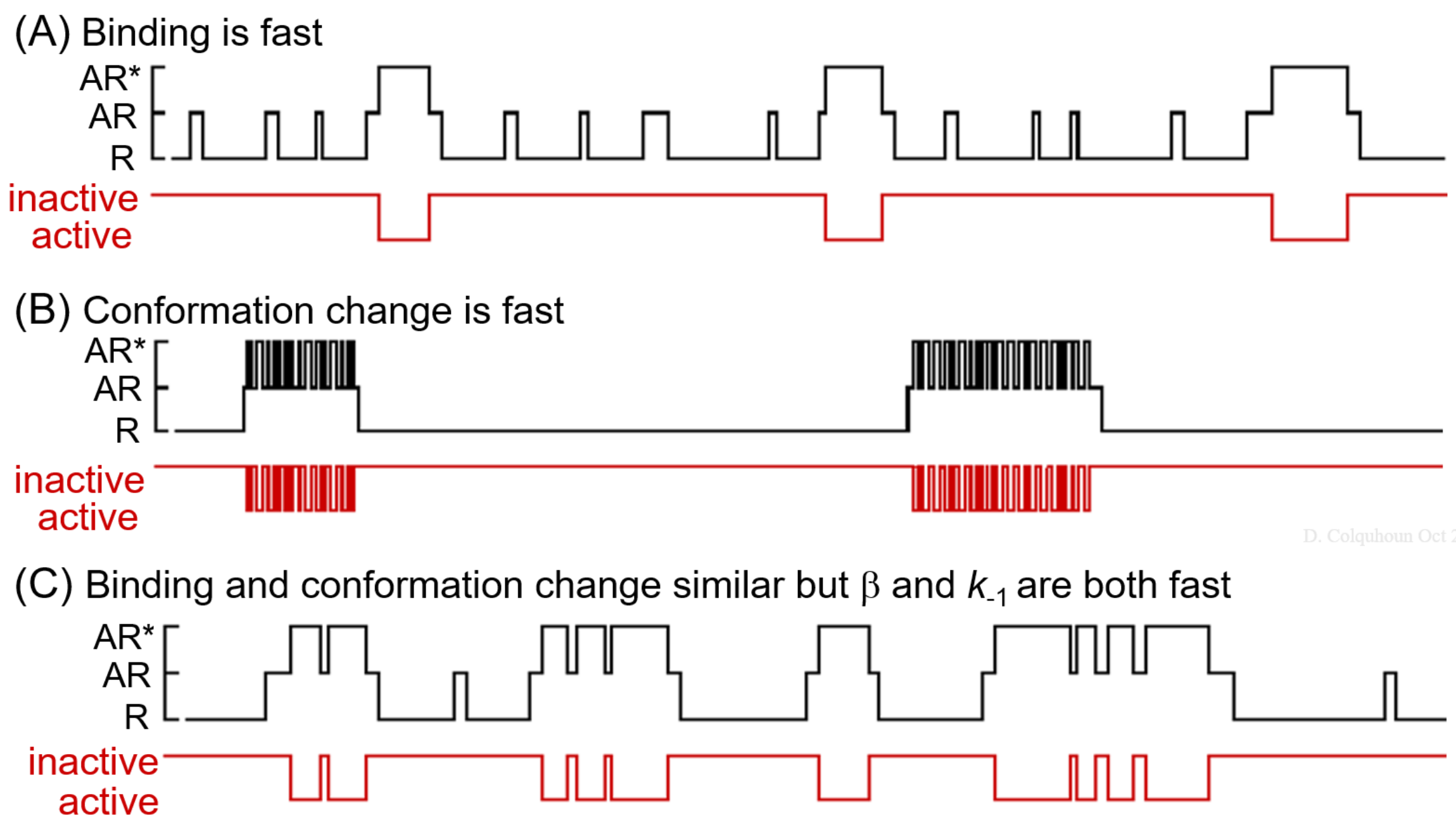

**Figure 1.** An illustration, at the single molecule level, of three cases of the del Castillo-Katz mechanism in which relaxation is well-approximated by a single exponential term. The black lines show typical transitions of the receptor between its three states. The red lines show the corresponding transitions between inactive (*i.e.*, closed, in the case of an ion channel) and active (*i.e.*, open, in the case of an ion channel) states of the receptor. (A) When binding is fast relative to the conformation change, there will be many brief bindings that don't produce an activation. After an activation does occur, the receptor will probably return to the resting state, so activations/openings mostly occur singly, *i.e.*, once per occupancy. (B) When the conformation change is fast relative to binding, every binding will result in activation and after the return to the intermediate, AR, state, the receptor is very likely to activate again many times before returning to the vacant state. (C) In the case where dissociation and activation are both fast, the time spent in the intermediate, AR, state will be low. When an activation occurs and the receptor returns to the intermediate state, the chances will be similar that it dissociates or that it reactivates. Activations will occur in bursts separated by short sojourns into the inactive states. The duration of the bursts will be very similar to the duration of ligand binding in this case.

In their analysis of acetylcholine-evoked current noise at the neuromuscular end-plate, Anderson & Stevens [29] postulated that nicotinic receptors could be described by the 'fast binding' approximation, in an attempt to explain why the decay of synaptic currents could usually be described well by a single exponential. Subsequent work showed that the 'low intermediate concentration' case was a much more likely explanation for this observation [5] [30].

### 3.2.1. The fast binding approximation

When the first step in the reaction, ligand binding, is much faster than the second step, isomerisation to the active state ($k_{+1} + k_{-1} \gg \alpha + \beta$), the binding step will, to a good approximation, remain at equilibrium throughout the relaxation. This means that the first two states in Scheme (10) can be pooled into a single state, so Scheme (10) can be written thus

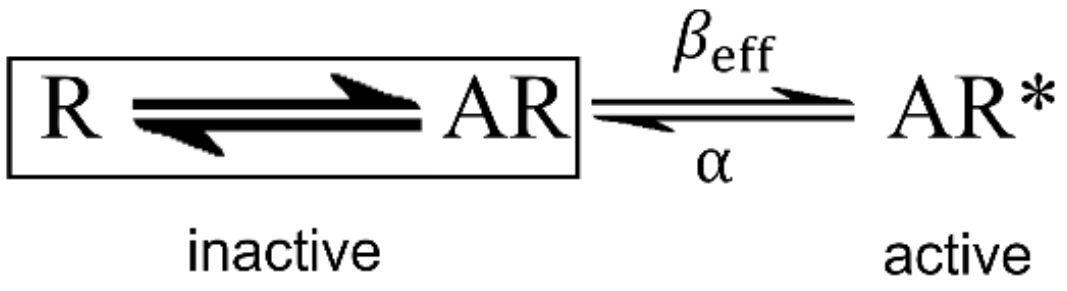

(32)

The assumption of fast binding means that the inactive states, R and AR, can be treated as a single state for kinetic purposes. To indicate this, they are enclosed in a box in Scheme (32). With this approximation, the system is similar to the two-state model in Scheme (1). The rate constant for activation must be altered because only bound channels, AR, can activate, so the effective activation rate constant is

$$\beta_{\mathrm{eff}} = \beta \frac{x_{\mathrm{A}}}{x_{\mathrm{A}} + K_{\mathrm{A}}} = \beta \frac{c_{\mathrm{A}}}{1 + c_{\mathrm{A}}} \tag{33}$$

The true activation rate constant, $\beta$, is multiplied by the equilibrium fraction of inactive receptors that are in the AR state. This is equivalent to the fraction of time for which a single inactive receptor, R or AR, is in the bound state, AR, and so capable of isomerising to the active state, AR*.

The behaviour of this mechanism at the level of a single molecule is shown in Figure 1A. As binding is fast, it is more likely that the ligand-receptor complex will dissociate rather than isomerise into the active conformation, so many ligand-receptor complexes will fail to activate. When an activation does occur, it's more likely to be followed by dissociation than by re-activation. The mean lifetime of the state AR is short, given by $1/(k_{-1} + \beta)$, and the probability that a receptor in the AR state will dissociate is $k_{-1}/(k_{-1} + \beta)$, which is large in this case. Therefore, most ligand-receptor complexes will produce only one activation before dissociating.

By analogy with the two-state mechanism, equilibration will follow a single exponential time constant, as in Eq. (4), with macroscopic (observed) rate constant being the sum of the two separate microscopic rate constants, as in Eq. (2), thus

$$\lambda = \alpha + \beta_{\mathrm{eff}} \equiv k_{\mathrm{on}} \tag{34}$$

This is the observed onset rate constant, $k_{on}$. The offset rate constant, $k_{\mathrm{off}}$, is

$$k_{\mathrm{off}} = \alpha \tag{35}$$

because the effective activation rate is zero when the free ligand concentration is zero.

To find the result of treating this special case of the dCK mechanism as though it were a simple binding reaction, we can substitute these expressions for $k_{\mathrm{on}}$ and $k_{\mathrm{off}}$ into Eq. (9). This gives the estimate for the dissociation equilibrium constant for binding as

$$K_{\mathrm{app}} \equiv \frac{k_{\mathrm{off}} x_{\mathrm{A}}}{k_{\mathrm{on}} - k_{\mathrm{off}}} = \frac{\alpha x_{\mathrm{A}}}{\beta_{\mathrm{eff}}} = \frac{K_{\mathrm{A}}(1 + c_{\mathrm{A}})}{E} \tag{36}$$

So, in this particular case, the standard way to estimate the equilibrium constant for binding from measured rates of binding fails to give either the microscopic equilibrium constant, $K_\mathrm{A} = k_{-1}/k_{+1}$, or the macroscopic equilibrium constant, $K_\mathrm{eff} = K_\mathrm{A}/(1+E)$. It is particularly unsatisfactory because the estimate of the equilibrium constant depends on the concentration of agonist used for the onset measurement.

But, in the case of fast binding, it may be possible to get an estimate of the microscopic equilibrium constant, $K_\mathrm{A}$, from the amplitude of the fast component. This is because the amplitude of the fast component at $t = 0$ provides the occupancy *before* any significant isomerisation into the active state, and, if binding is fast enough, this will be the equilibrium occupancy. Therefore, $K_\mathrm{A}$ can be estimated by rearranging Eq. (5), as

$$\frac{x_\mathrm{A}(1-p)}{p} \tag{37}$$

where $p$ is the amplitude of the fast component. An example of this calculation is given below, in Section 3.3.2. We aren't aware of any cases in which it has been possible to measure the amplitude of the predicted initial jump in occupancy. Numerical examples of cases in which binding is fast are given in Section 3.3.2 and 3.3.3, and this method works in principle.

### 3.2.2. The fast conformation change approximation

If the isomerisation reaction is very fast compared with rate of binding ($\alpha + \beta \gg k_{+1} + k_{-1}$), then the two bound species, inactive and active, can be treated as though they are permanently at equilibrium, so they can be considered as a single state for kinetic purposes, as shown by the box enclosing them in Scheme (38).

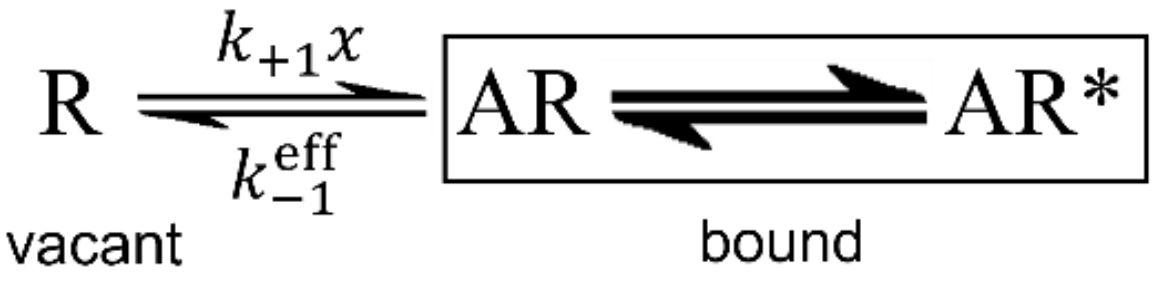

(38)

The behaviour of this mechanism at the level of a single molecule is shown in Figure 1B. The effective rate of dissociation, $k_{-1}^{\mathrm{eff}}$, will be the actual rate, $k_{-1}$, multiplied by the equilibrium fraction of bound receptors that are in the AR state, the only state from which dissociation is possible. Thus

$$k_{-1}^{\mathrm{eff}} = \frac{k_{-1}}{1+E} \tag{39}$$

We now have only two states, vacant and bound, so the macroscopic rate of equilibration which we predict to be observed will be

$$\lambda = k_{-1}^{\mathrm{eff}} + k_{+1}x_\mathrm{A} \equiv k_\mathrm{on} \tag{40}$$

For the offset of binding, when $x_\mathrm{A} = 0$,

$$k_{-1}^{\text{eff}} = k_{\text{off}} \tag{41}$$

Substitution of these expressions for $k_{\text{on}}$ and $k_{\text{off}}$ into Eq. (9) gives an apparent equilibrium constant of

$$K_{\text{app}} = \frac{k_{\text{off}} x_{\text{A}}}{k_{\text{on}} - k_{\text{off}}} = \frac{k_{-1}^{\text{eff}}\, x_{\text{A}}}{k_{+1} x_{\text{A}}} = \frac{K_{\text{A}}}{1+E} \tag{42}$$

In this case, the rates of binding (or of response) give the effective equilibrium constant, Eq. (31), which depends on both $K_{\text{A}}$, the actual microscopic equilibrium constant for binding, and the agonist's efficacy, $E$. A numerical example is given in Section 3.3.4. It's unlikely that the fast conformation change case is a good approximation for ligand-gated ion channels because channel openings don't behave as depicted in Figure 1B [5] [30] [13] [14].

### 3.2.3. The low intermediate concentration approximation

There is a third, less well-known, case in which only one of the two exponential components will be detectable. That is the case when the concentration of the intermediate species, AR, is always low; in single molecule terms this means that the mean lifetime of AR is short. This will be the case when the rate constants that lead away from this state are both fast, even if the binding reaction and conformation change have a similar overall rate. These rate constants are $k_{-1}$ for dissociation and $\beta$ for activation, as indicated by the thick arrows in Scheme (43).

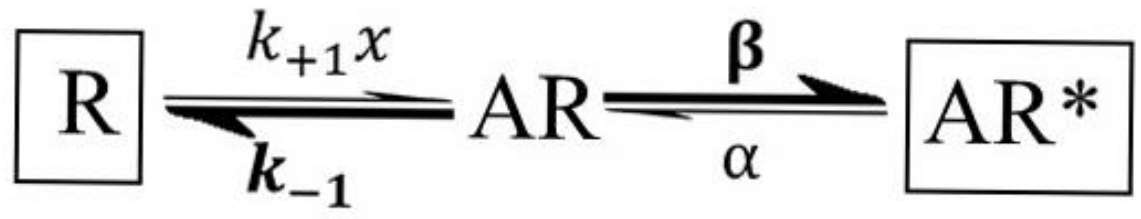

(43)

If both of these fast rate constants are comparable then an activation will usually consist of a few sojourns in the active state, AR*, separated by very short times in the intermediate state, AR, as illustrated in Figure 1C. The length of such bursts of activations, will be nearly exponentially distributed. This conclusion follows from the fact that the sum of a random (geometrically-distributed) number of exponentially distributed variables is itself exponentially distributed – see Section 9.3 in ref. [25].

The mean duration of a sojourn in AR is $1/(\beta + k_{-1})$, and the mean number of activations per burst is $1 + (\beta/k_{-1})$ [5] [6] – the faster the opening rate, $\beta$, is relative to the dissociation rate, $k_{-1}$, the more often the receptor, when it's in state AR, will re-activate rather than dissociate. The mean length of each activation is $1/\alpha$ so the mean time spent in the active state per burst, which we can define as $1/\alpha_{\text{eff}}$, is

$$\frac{1}{\alpha_{\text{eff}}} \equiv \frac{1}{\alpha}\left(1 + \frac{\beta}{k_{-1}}\right) \tag{44}$$

And, because the shut times within a burst of openings are short, this is close to the mean length of a burst of activations. The exact expressions for burst length are given in [6].

We next need the mean length of the shut time between bursts of activations. During the intervals between bursts there must be at least one sojourn in the vacant state, R, but there may be several (very short) occupancies in AR, followed by return to R. By analogy with bursts of openings, the mean number of sojourns in R will be $1 + k_{-1}/\beta$, – the faster the dissociation rate, $k_{-1}$, is relative to the opening rate, $\beta$, the more often the receptor, when it's in state AR, will dissociate rather than re-activate. And the mean length of each sojourn in R is $1/k_{+1}x_{\mathrm{A}}$, so the mean total time in R between bursts is

$$\frac{1}{k_{+1}x_{\mathrm{A}}}\left(1 + \frac{k_{-1}}{\beta}\right) \tag{45}$$

This will be the mean length of the time between bursts of activations if we neglect the short times spent in AR. So, as in the two-state case, Eq. (2), the macroscopic (observed) rate constant for re-equilibration can be written, to a good approximation, as

$$\lambda = \frac{1}{\text{mean lifetime between bursts of activations}} + \frac{1}{\text{mean lifetime of bursts of activations}}$$

$$\approx \frac{k_{+1}x_{\mathrm{A}}}{\left(1 + \dfrac{k_{-1}}{\beta}\right)} + \frac{\alpha}{\left(1 + \dfrac{\beta}{k_{-1}}\right)} = \frac{\beta k_{+1}x_{\mathrm{A}} + \alpha k_{-1}}{\beta + k_{-1}} \equiv k_{\mathrm{on}} \tag{46}$$

This gives, to a good approximation, the onset rate constant that should be observed. Note that, for any Markov process, the time constants for re-equilibration are the same for all states of the system, and for any linear combination of states, such as ligand binding or current flow. Therefore, it doesn't matter that this argument has been presented in terms of receptor activation: the macroscopic time constant(s) are the same for ligand binding and for activation, only their amplitudes differ.

For the offset of binding, we set the free concentration to zero, $x_{\mathrm{A}} = 0$, in Eq. (46), so

$$k_{\mathrm{off}} = \frac{\alpha k_{-1}}{\beta + k_{-1}} \tag{47}$$

We can now substitute $k_{\mathrm{on}}$ and $k_{\mathrm{off}}$, from Eq. (46) and (47), into Eq. (9) to find the estimate of the equilibrium dissociation constant from the observed rates of binding onset and offset. The result is

$$K_{\mathrm{app}} = \frac{k_{\mathrm{off}}x_{\mathrm{A}}}{k_{\mathrm{on}} - k_{\mathrm{off}}} = \frac{\alpha k_{-1}}{\beta k_{+1}} = \frac{K_{\mathrm{A}}}{E} \tag{48}$$

So once again, the apparent equilibrium constant for ligand binding, estimated from the rates of the onset and offset of binding, reflects both of the equilibrium constants in the underlying mechanism, the affinity for binding, $K_{\mathrm{A}}$, and the ability to change conformation, the efficacy, $E$.

The assumption that the activation rate, $\beta$, is fast implies that the efficacy, $E = \beta/\alpha$, is large (if $\alpha$ was also fast, we'd have the 'fast conformation change' case, dealt with in the previous section). And, if $E$ is large then both of these cases give a result that is close to the macroscopic binding constant, $K_{\mathrm{eff}}$, defined in Eq. (31). A numerical example follows in Section 3.3.5.

## 3.3. Numerical examples of the del Castillo-Katz mechanism

Examples of the time course of ligand binding and receptor activation are calculated with an R program, `CK-kinetics.R` (see the electronic supplementary material), a more limited version of an older DOS program, SCALCS [31,32]. Another version of SCALCS exists in Python, written by Remis Lape [33]. Each run of the program produces three graphs: the time course of the fraction of receptors in each state, the time course of ligand binding, and the time course of activation/opening. The last two of these also show the individual exponential components (dashed orange lines). The concentration of ligand is supposed to jump from 0 to $x_{\mathrm{A}}$ at time zero and then jump back to zero a few milliseconds later – the duration of the application is defined in the program as `tpulse`. The R program also produces a text file with details of the results. For details about how to run the R program, see `readme.pdf`, in the electronic supplementary materials.

The values of the rate constants used in the numerical examples are based roughly on numbers that would be sensible for ion channels. Surface plasmon resonance has a temporal resolution of about $100$ ms (Biacore 2000 Instrument Handbook), far slower than the resolution of about $20\ \mu\mathrm{s}$ that can be achieved in single ion channel recordings or $100\ \mu\mathrm{s}$ that can be achieved for concentration jumps on outside-out patches [15]. Therefore, for ligand binding experiments, all the rates in the examples should be divided by a factor of 10 to 1000 to match the rates observed in binding experiments.

### 3.3.1. A case in which two components are detectable

First consider a case in which neither of the exponential components predominates. An example is shown in Figure 2. Details are given in the printout file, "`two components v5.7.txt`", in electronic supplementary material.

In this example, the activation/opening rate constant is $\beta = 1000\ \mathrm{s}^{-1}$, the deactivation/shutting rate constant is $\alpha = 200\ \mathrm{s}^{-1}$, the dissociation rate constant is $k_{-1} = 1000\ \mathrm{s}^{-1}$, the association rate constant is $k_{+1} = 5 \times 10^{8}\ \mathrm{M}^{-1}\mathrm{s}^{-1}$ and $x_{\mathrm{A}} = 2.5\ \mu\mathrm{M}$. So, the pseudo first order association rate is $k_{+1}x_{\mathrm{A}} = 1250\ \mathrm{s}^{-1}$. These are the numbers that go into the $\boldsymbol{Q}$ matrix, defined in Eq. (20). These rate constants also define the microscopic dissociation equilibrium constant for binding, $K_{\mathrm{A}} = k_{-1}/k_{+1} = 2\ \mu\mathrm{M}$, the equilibrium constant for the isomerisation between inactive state, AR, to the active state, AR*, $E = \beta/\alpha = 5$ and the effective (macroscopic) binding constant, $K_{\mathrm{eff}} =$

$K_A/(1+E) = 0.333\ \mu$M. The time course of onset of ligand binding, given in the printout file, is

$$p_{\text{bound}}(t) = 0.882 - 0.321 \exp\left(-\frac{t}{0.350}\right) - 0.562 \exp\left(-\frac{t}{1.68}\right) \qquad (49)$$

Figure 2B shows this time course of ligand binding. The fraction of binding sites that are occupied rises from zero at $t = 0$ towards its equilibrium value of 0.882. The time course of the rise in occupancy is described by the sum of two exponential curves; the faster one has a time constant of 0.350 ms and the slower one has a time constant of 1.68 ms. The amplitude of the faster component is more than half that of the slower ($0.562/0.321 \approx 1.75$). It's clear from Figure 2B that the onset of binding starts too fast for the slower component alone to be a good fit. Figure 2A shows the time courses of the occupancies of all three states. It's clear from this that the reason for the initial fast binding is the fast rise of the concentration of the intermediate state, AR (red curve).

The length of the ligand application was 5 ms in Figure 2. At $t = 5$ ms, the free ligand concentration is reduced to $x_A = 0$. The occupancies at $t = 5$ ms give the initial occupancies, $\boldsymbol{p}(0)$, for the separate calculation of the time course of ligand dissociation. The time course of dissociation of bound ligand (the offset step) is given in the printout as

$$p_{\text{bound}}(t) = 0.0422 \exp\left(-\frac{t}{0.475}\right) + 0.811 \exp\left(-\frac{t}{10.5}\right) \qquad (50)$$

The equilibrium fraction bound at zero concentration is zero. As usual, offset is slower than onset, with the two time constants for offset being 0.475 ms and 10.5 ms. But the relative amplitude of the slower component is much bigger ($0.811/0.0422 \approx 19$-fold greater), so it's unlikely that the time course of offset could be distinguished experimentally from a single exponential with a time constant of 10.5 ms (as indicated by the dashed orange line in Figure 2B). This is in contrast with the onset curve where the corresponding ratio of amplitudes was only 1.75, so both components should be detectable.

The time course of receptor activation is shown in Figure 2C. It shows a sigmoid rise – a delay – during onset. It's clear from the time courses of state occupancies in Figure 2A that this is a result of the initial fast buildup of the bound but inactive intermediate state, AR (red line).

For the three-state dCK mechanism, there is only one active state, so the activation time course in Figure 2C is the same as the time course of state 1, AR*, shown as the blue curve in Figure 2A. Therefore, the activation curve will not be shown separately in subsequent examples of the dCK mechanism.

Given that there are clearly two components of comparable amplitude, it isn't possible to calculate the binding constant from the onset and offset rates.

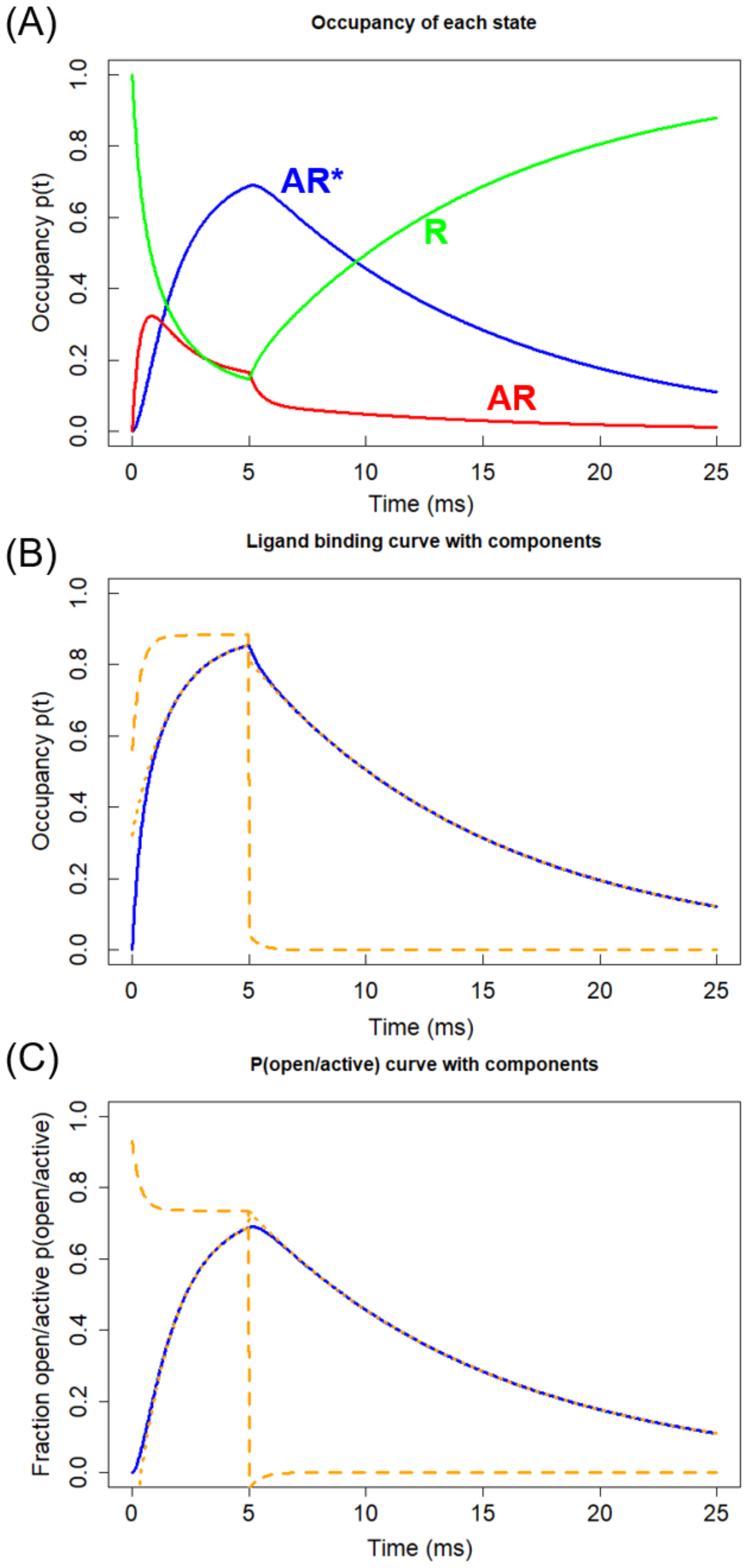

**Figure 2.** A case in which two components are detectable. (A) The time course of the occupancy of each of the three states in the dCK mechanism during a 5 ms pulse of agonist ($2.5\ \mu\mathrm{M}$). (B) The time course of ligand binding during a 5 ms pulse of agonist. This reflects the occupancy of the AR and AR* states. (C) The time course of receptor activation – that is, the occupancy of the AR* state. In (B) and (C), the observed time courses of binding and activation are shown in blue, with the components shown in orange. For all binding curves, the equilibrium occupancy is added to the components of onset (in orange) to ensure superimposition with the time course of binding (in blue) when the other components are negligible. For details of this example, see "`two components v5.7.txt`" in the electronic supplementary material.

### 3.3.2. A case of fast binding

The case of fast binding was discussed in Section 3.2.1, illustrated in Figure 1A, and it's exemplified in Figure 3. The details of this example are given in the printout, "`fast binding v5.7.txt`" (see the electronic supplementary material).

In this case the dissociation rate is fast, $k_{-1} = 100{,}000\ s^{-1}$, and $k_{+1} = 1 \times 10^{9}\ \mathrm{M}^{-1}\mathrm{s}^{-1}$ and $x_{\mathrm{A}} = 100\ \mu\mathrm{M}$ so the association rate is $k_{+1}x_{\mathrm{A}} = 100{,}000\ \mathrm{s}^{-1}$ – also very fast. The isomerisation rates are much slower, $\beta = 5{,}000\ \mathrm{s}^{-1}$ and $\alpha = 500\ \mathrm{s}^{-1}$. These values define the equilibrium constants $K_{\mathrm{A}} = 100\ \mu\mathrm{M}$, $E = 10$ and $K_{\mathrm{eff}} = 9.09\ \mu\mathrm{M}$. Figure 3A shows that binding, moving from state R to state AR, occurs almost instantaneously (on the time scale of the measurements), then the concentration of AR falls slowly as AR isomerises to the active state, AR*. Correspondingly, the ligand binding curve, Figure 3B, shows a near instantaneous jump when the ligand is applied as state AR becomes occupied, and then rises more slowly as the active state is formed. During the offset of ligand binding, a corresponding pattern is evident.

The appearance of the ligand binding curve in this concentration jump experiment will be familiar to electrophysiologists who have done a voltage jump experiment. The initial jump in current defines the instantaneous current-voltage relationship. This reflects the properties of individual ion channels because it measures the change in current flow in response to a step change in voltage through a number of open channels that hasn't had time to change in response to the change in voltage.

The printout file gives the time course of the onset of ligand binding as

$$p_{\mathrm{bound}}(t) = \ 0.917 - \ 0.487 \exp\left(-\frac{t}{0.00494}\right) - \ 0.429 \exp\left(-\frac{t}{0.338}\right) \qquad (51)$$

The initial fast jump, with a time constant of $4.9\ \mu\mathrm{s}$, would be too fast for its rate to be measurable, even in the case of ligand-gated ion channels. The slower component, with a time constant of 0.338 ms, is what would be measured and it is very close to a single exponential, as shown by the superposition of the orange dashed line on the blue line in Figure 3B.

Note that the offset time constant, 2.1 ms, is close to the mean length of a burst of activations. In this case, activations mostly come singly (the mean number per burst is 1.05) because, once in the AR state, dissociation is much more likely than re-activation. So the mean active/open time per burst is $(1/\alpha)\big(1 + (\beta/k_{-1)})\big) = 2$ ms x 1.05 = 2.10 ms. This is expected on the basis of the general rule that the time constants for macroscopic relaxation after a jump to zero concentration are the same as the time constants for the distribution of the single channel burst length [34].

The predominant time constant for onset, 0.338 ms, is quite close to its theoretical value for very fast binding, given by Eq. (34), of 0.333 ms. And for the offset, the predominant time constant, 2.1 ms, is close to the value from Eq. (35), namely $1/\alpha = 2$ ms.

The amplitude of the initial jump in occupancy gives a good estimate of the microscopic binding constant, $K_\mathrm{A} = 100\ \mu\mathrm{M}$. Substituting the amplitude of the fast jump, 0.487, into Eq. (37) gives $105\ \mu\mathrm{M}$, a fairly good estimate of $K_\mathrm{A}$.

The estimate of affinity provided by the observed onset and offset rates is not good. It's 19.1 $\mu\mathrm{M}$, which isn't close to either $K_\mathrm{A} = 100\ \mu\mathrm{M}$ or $K_\mathrm{eff} = 9.09\ \mu\mathrm{M}$, though it is close to the value expected for very fast binding from Eq. (36), given by

$$\frac{K_\mathrm{A}(1 + c_\mathrm{A})}{E} = 20\ \mu\mathrm{M} \qquad (52)$$

This estimate is unsatisfactory because it depends on ligand concentration and because it doesn't allow separate estimation of either affinity, $K_\mathrm{A}$, or of efficacy, $E$.

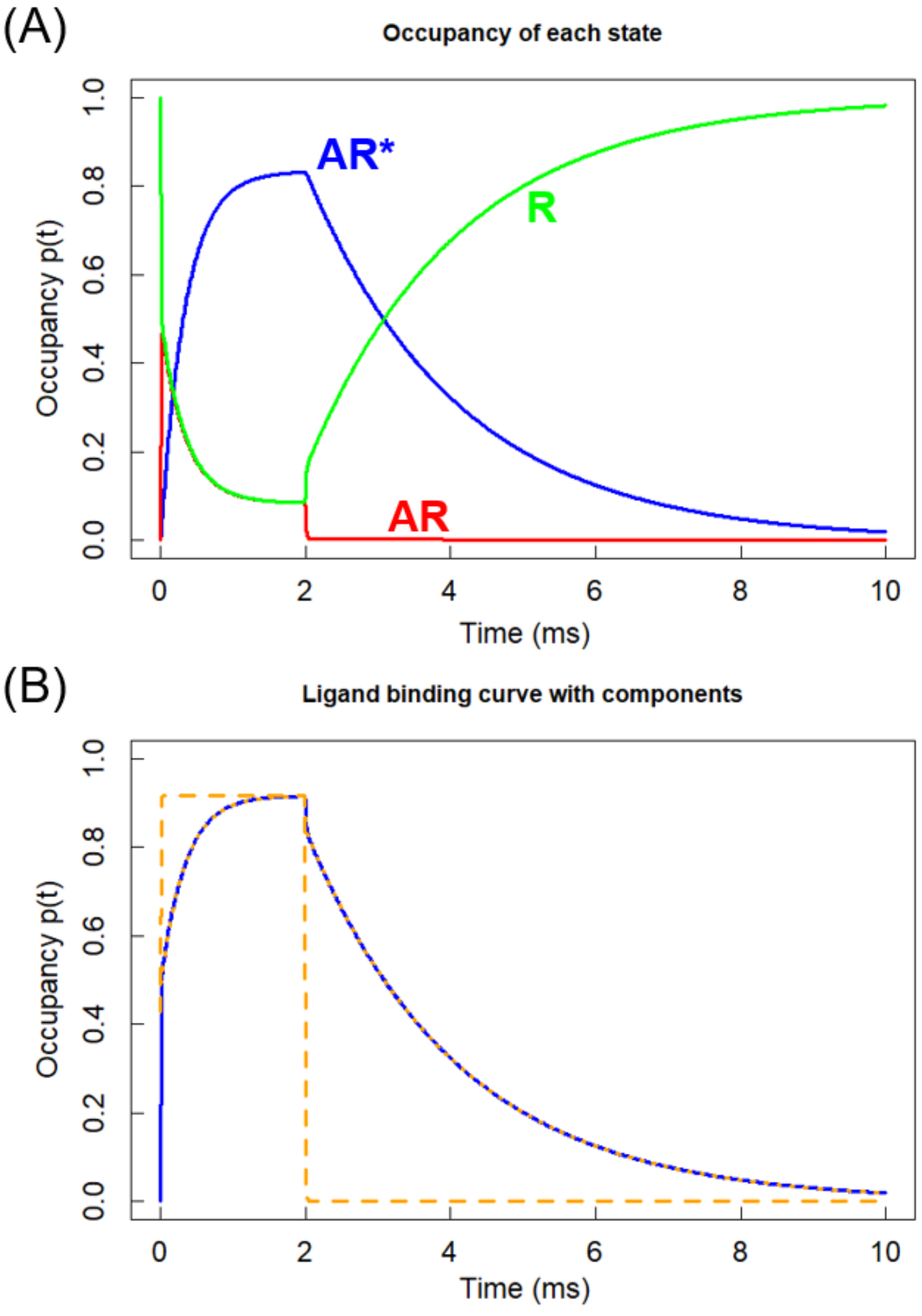

**Figure 3.** A case in which binding is fast. (A) The time course of the occupancy of each of the three states in the dCK mechanism during a 2 ms pulse of agonist $(100\,\mu\mathrm{M})$. Notice the near-instantaneous rise in the occupancy of the AR state, reflecting very rapid ligand binding, followed by a slower transition to the AR* state. States R and AR stay essentially in equilibrium during the onset. (B) The time course of ligand binding during a 2 ms pulse of agonist. The observed time course of binding is shown in blue, with the components shown in orange. Plots are produced by `CK-kinetics.R`, with details in the printout, "`fast binding v5.7.txt`" (see the electronic supplementary material).

#### 3.3.3. A case of fast-*ish* binding

Another example of fast binding is shown in Figure 4, in which the rate constants for binding have been reduced 10-fold from the fast binding example, above. The numbers are in the printout file "`fastish binding.txt`" (see the electronic supplementary material). In this case the dissociation rate is $k_{-1} = 10{,}000\ s^{-1}$, and $k_{+1} = 1 \times 10^{9}\ \mathrm{M}^{-1}\mathrm{s}^{-1}$ and $x_{\mathrm{A}} = 10\ \mu\mathrm{M}$ so the association rate is $k_{+1}x_{\mathrm{A}} = 10{,}000\ \mathrm{s}^{-1}$.

The isomerisation rates are much slower, $\beta = 1{,}000\ \mathrm{s}^{-1}$ and $\alpha = 100\ \mathrm{s}^{-1}$. These values define the equilibrium constants $K_\mathrm{A} = 10\ \mu\mathrm{M}$, $E = 10$ and $K_\mathrm{eff} = 0.909\ \mu\mathrm{M}$.

As in the previous example, there is a fast accumulation of the AR state before isomerisation into the AR* state (Figure 4A). The ligand binding curve in Figure 4B still rises rapidly at first, but the faster time constant for onset, 0.0487 ms, is slower than the previous example so the initial fast rise in the onset of binding no longer looks ‘instantaneous’, and both components are visible. Both components are also clearly visible in the offset of binding. If the rates of the slow components are used to estimate the equilibrium constant, the result is $1.84\ \mu\mathrm{M}$, which isn’t close to either $K_\mathrm{A} = 10\ \mu\mathrm{M}$, or to $K_\mathrm{eff} = 0.909\ \mu\mathrm{M}$, as expected from Eq. (36).

Despite the fact that the initial jump in ligand binding (Figure 4B) doesn’t look ‘instantaneous’ in this case, the estimate of the microscopic affinity constant from the amplitude of the fast component of onset, $11.1\ \mu\mathrm{M}$, is not far from the true value, $K_\mathrm{A} = 10\ \mu\mathrm{M}$.

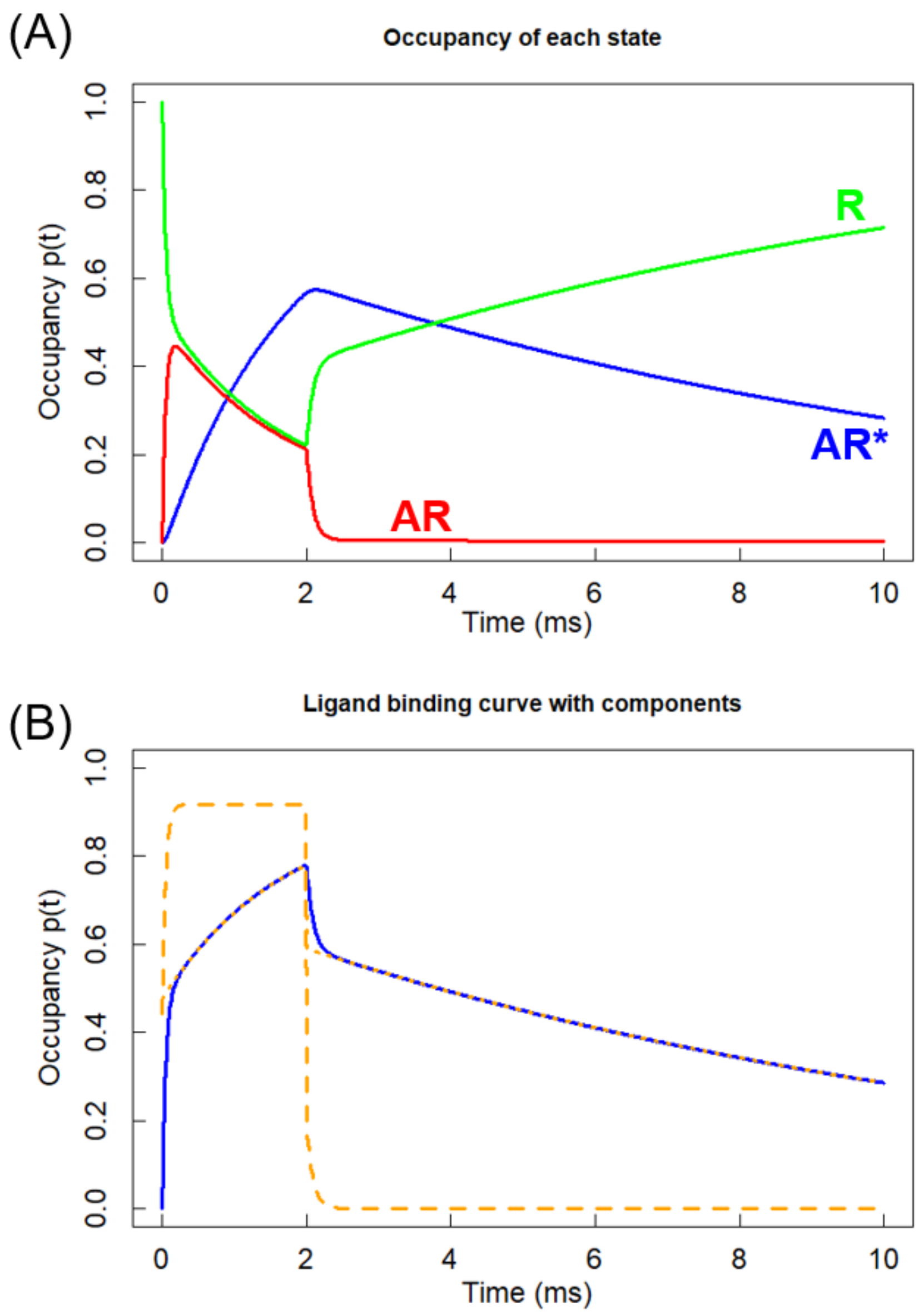

**Figure 4.** Another example of a case in which binding is fast. (A) The time course of the occupancy of each of the three states in the dCK mechanism during a 2 ms pulse of agonist (10 μM). As in Figure 3, there is a fast rise in the occupancy of the AR state, reflecting rapid ligand binding, followed by a slower transition to the AR* state. (B) The time course of ligand binding during a 2 ms pulse of agonist. The observed time course of binding is shown in blue, with the components shown in orange. Plots are produced by `CK-kinetics.R`. For details, see the printout file "`fastish binding v5.7.txt`" in the electronic supplementary material.

### 3.3.4. A case of fast conformation change

The second case in which onset and offset are close to following a single exponential time course is the case when the conformation change step is much faster than ligand binding. This was described in Section 3.2.2, and the results of this numerical example are shown in Figure 5 and in the printout file "`fast conformation v5.7.txt`" which can be found in the electronic supplementary material. The single molecule behaviour in this case is illustrated in Figure 1B.

For this example, we set the isomerisation rates as $\beta = 50{,}000\ \mathrm{s}^{-1}$ and $\alpha = 5000\ \mathrm{s}^{-1}$, faster than the dissociation rate $k_{-1} = 1000\ \mathrm{s}^{-1}$. Here, $k_{+1} = 1 \times 10^{8}\ \mathrm{M}^{-1}\mathrm{s}^{-1}$, with concentration $x_{\mathrm{A}} = 10\ \mu\mathrm{M}$, which gives the association rate $k_{+1}x_{\mathrm{A}} = 1000\ \mathrm{s}^{-1}$, much slower than the rates of conformation change. These values define the equilibrium constants $K_{\mathrm{A}} = 10\ \mu\mathrm{M}$, $E = 10$ and $K_{\mathrm{eff}} = 0.909\ \mu\mathrm{M}$.

The onset of ligand binding is given by

$$p_{\mathrm{bound}}(t) = 0.917 - 0.0003 \exp\left(-\frac{t}{0.0179}\right) - 0.916 \exp\left(-\frac{t}{0.932}\right) \qquad (53)$$

The amplitude of the fast component is tiny, so the equilibrium occupancy, 91.7%, is approached on a time course that is very close to a single exponential with a time constant of 0.932 ms. This is illustrated by the superimposition of the orange dashed line on the blue time course in Figure 5B.

At equilibrium, once the receptor is in the AR state, it is very likely to reactivate rather than dissociate. Here, the probability of reactivation from the AR state is $\beta/(\beta + k_{-1}) = 0.98$. The mean number of activations in the bursts of activations illustrated in Figure 1B is $1 + (\beta/k_{-1}) = 51$. The mean length of an activation is $1/\alpha = 0.2\ \mathrm{ms}$, so the mean active/open time per burst is 10.2 ms. The 51 activations are separated by 50 brief sojourns in the AR state, each of mean length $1/(\beta + k_{-1}) = 19.6\ \mu\mathrm{s}$. These add up to 0.98 ms so the mean total length of a burst of activations is 10.2 + 0.98 = 11.18 ms. This is very close to the predominant offset time constant in this example, 11.2 ms, as is expected when the gaps within bursts of activations (sojourns in the AR state) are brief. This is expected on the basis of the general rule that the time constants for macroscopic relaxation after a jump to zero concentration are the same as the time constants for the distribution of the single channel burst length [34]. Exact expressions for quantities like burst lengths are given in [25].

The predominant time constant for onset, 0.932, is close to the theoretical limiting value for the fast conformation change approximation given by Eq. (40), which is 0.917 ms. And the predominant time constant for offset, 11.2, ms is close to the theoretical limiting value given by Eq. (41) as 11.0 ms. Therefore, it's not surprising that the estimate of the equilibrium constant calculated from the predominant onset and offset rates is $0.9094\ \mu\mathrm{M}$, very close indeed to $K_{\mathrm{eff}} = K_{\mathrm{A}}/(1 + E) = 0.9091\ \mu\mathrm{M}$. This is what's expected from Eq. (42), when the fast conformation change approximation is appropriate.

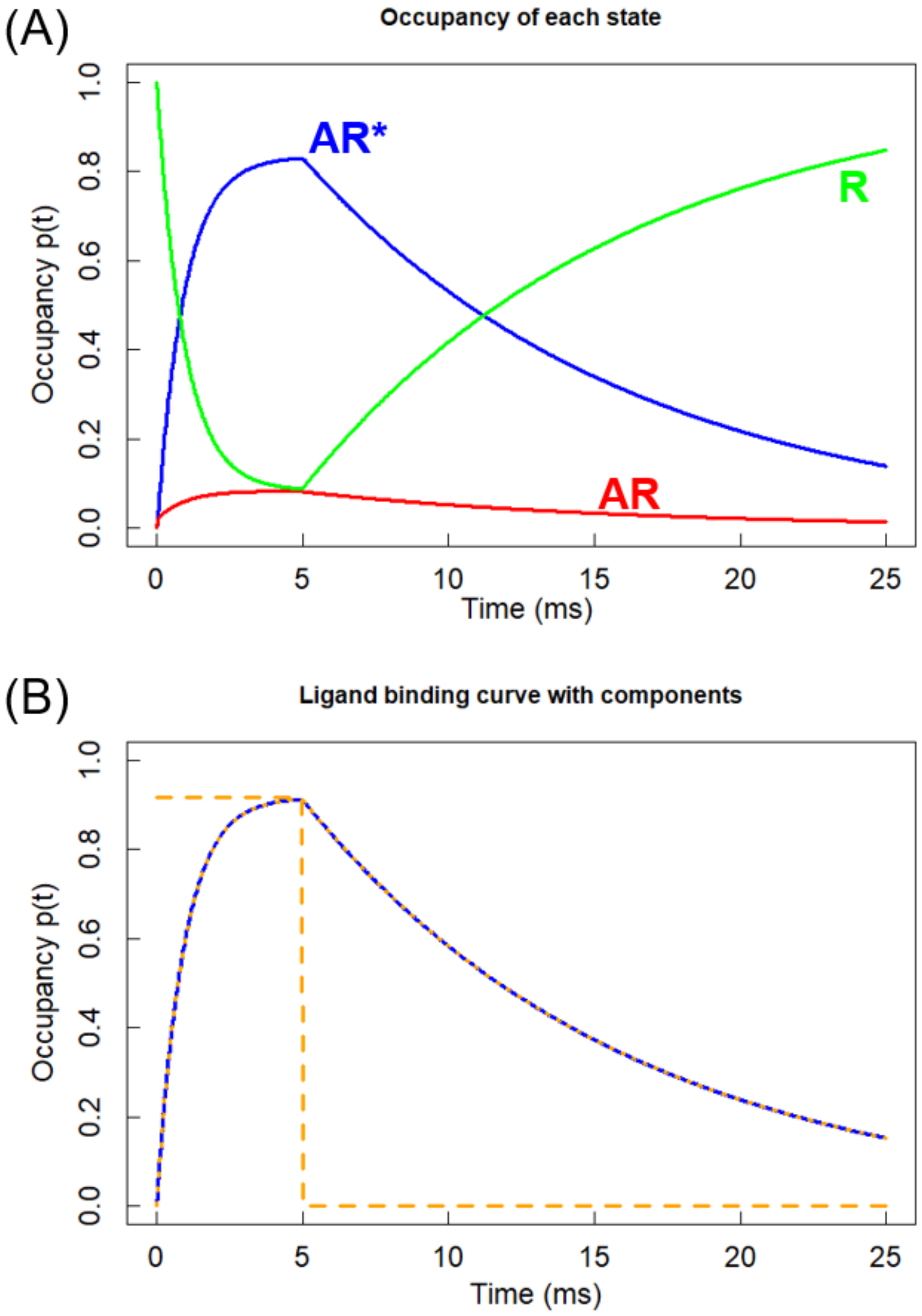

**Figure 5.** A case in which the conformation change is fast. (A) The time course of the occupancy of each of the three states in the dCK mechanism during a 5 ms pulse of agonist (10 μM). Unlike the examples of fast binding shown in Figures 3 and 4, there is no rapid accumulation of the AR state at the start of the pulse of agonist because of the fast isomerisation to the AR* state. (B) The time course of ligand binding during a 5 ms pulse of agonist. The observed time course of binding is shown in blue, with the components shown in orange. Plots are produced by `CK-kinetics.R`, For details, see the printout file, "`fast conformation v5.7.txt`", in the electronic supplementary material.

### 3.3.5. A case of low intermediate concentration

The third special case of the dCK mechanism is illustrated in Figure 1C and analysed in Section 3.2.3. The numbers for the numerical example, shown in Figure 6, are in the printout file, "`low intermediate v5.7.txt`", which is in the electronic supplementary material.

In this case, both the dissociation rate constant, $k_{-1}$, and the activation rate constant, $\beta$, are fast, so the lifetime of the intermediate state, AR, given by $1/(k_{-1}+\beta)$, is short and its concentration (or fractional occupancy) is always low. We set $k_{-1}=\beta=20{,}000\ \mathrm{s}^{-1}$. The association rate constant, $k_{+1}=5\times10^{8}\ \mathrm{M}^{-1}\mathrm{s}^{-1}$, and free ligand concentration, $x_{\mathrm{A}}=\ 10\ \mu\mathrm{M}$, give the association rate as $k_{+1}x_{\mathrm{A}}=\ 5000\ \mathrm{s}^{-1}$, and we set $\alpha=1000\ \mathrm{s}^{-1}$. These values define the equilibrium constants $K_{\mathrm{A}}=40\ \mu\mathrm{M}$, $E=20$ and $K_{\mathrm{eff}}=K_{\mathrm{A}}/(1+E)=1.90\ \mu\mathrm{M}$. The mean lifetime of the intermediate state, AR, is brief

$$\frac{10^{6}}{k_{-1}+\beta}=25\ \mu\mathrm{s} \tag{54}$$

Figure 6A shows that the concentration of the intermediate state, AR (red line), stays low throughout the application of the ligand.

The onset time course for ligand binding is described by

$$p_{\mathrm{bound}}(t)=0.840-0.0638\exp\left(-\frac{t}{0.0232}\right)-0.776\exp\left(-\frac{t}{0.345}\right) \tag{55}$$

The fast component is too small in amplitude, and too fast, to be detected, so ligand binding increases from zero to its equilibrium value, 84% occupancy, with a time course that's very close to a single exponential with the slower time constant, 0.345 ms, as shown in Figure 6B.

The offset of ligand binding is also very close to a single exponential, with a time constant of 2.03 ms. This is close to the mean length of a burst of activations/openings (see Figure 1C) which is 2.025 ms (on average there are 2 activations of 1 ms duration each per burst, separated by a brief sojourn in AR of 0.025 ms).

The predominant onset time constant, 0.345 ms, is close to the approximate onset time constant for this case, from Eq. (46), which gives 0.333 ms. The offset curve is similarly dominated by its slow component, with a time constant of 2.10 ms. This is again close to the approximate offset time constant for this case, given by Eq. (47) as 2.0 ms.

The estimate of the equilibrium constant calculated from the predominant onset and offset rates is $2.05\ \mu\mathrm{M}$, which is close to $K_{\mathrm{eff}}=K_{\mathrm{A}}/(1+E)=\ 1.90\ \mu\mathrm{M}$. This is what's expected from Eq. (48), when the low intermediate concentration approximation is appropriate and *E* is large. Eq. (48) gives the result in the low intermediate concentration case as $K_{\mathrm{A}}/E=2.0\ \mu\mathrm{M}$.

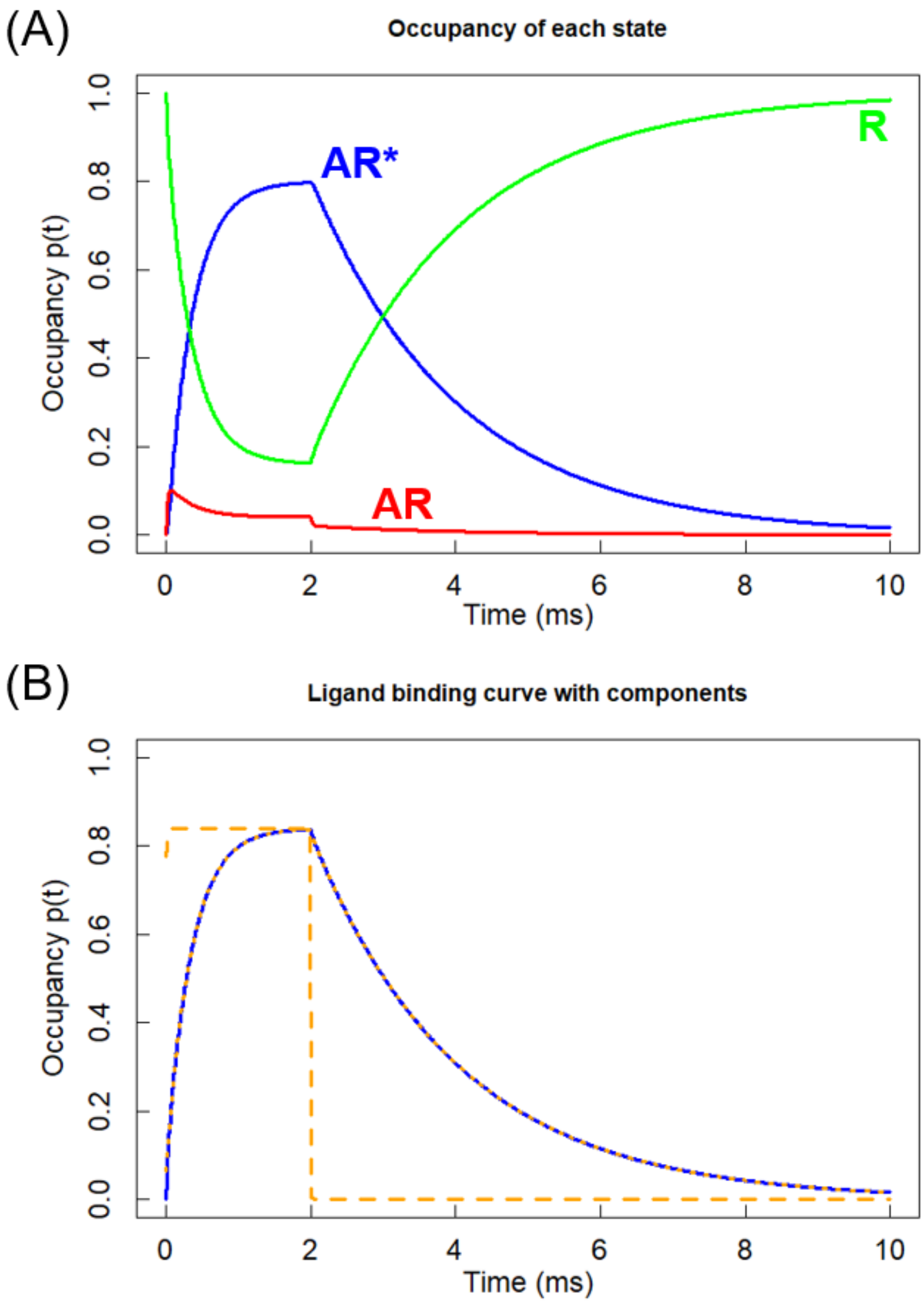

**Figure 6.** A case in which the concentration of the intermediate, AR, state is low. (A) The time course of the occupancy of each of the three states in the dCK mechanism during a 2 ms pulse of agonist (10 μM). (B) The time course of ligand binding during a 2 ms pulse of agonist. The observed time course of binding is shown in blue, with the components shown in orange. For details, see "`low intermediate v5.7`" in the electronic supplementary material.

## 4. A model with more than one binding site

Many receptors need to bind more than one ligand molecule to become active. A simple model of this sort was postulated for the nicotinic acetylcholine receptor by Colquhoun and Hawkes [5]. The model is shown in Scheme (56) and will be referred to as the CH82 mechanism. Neither the model, nor the numerical values of the rate constants, are intended to be descriptions of real receptors, and this model has since been superseded [13] [14].

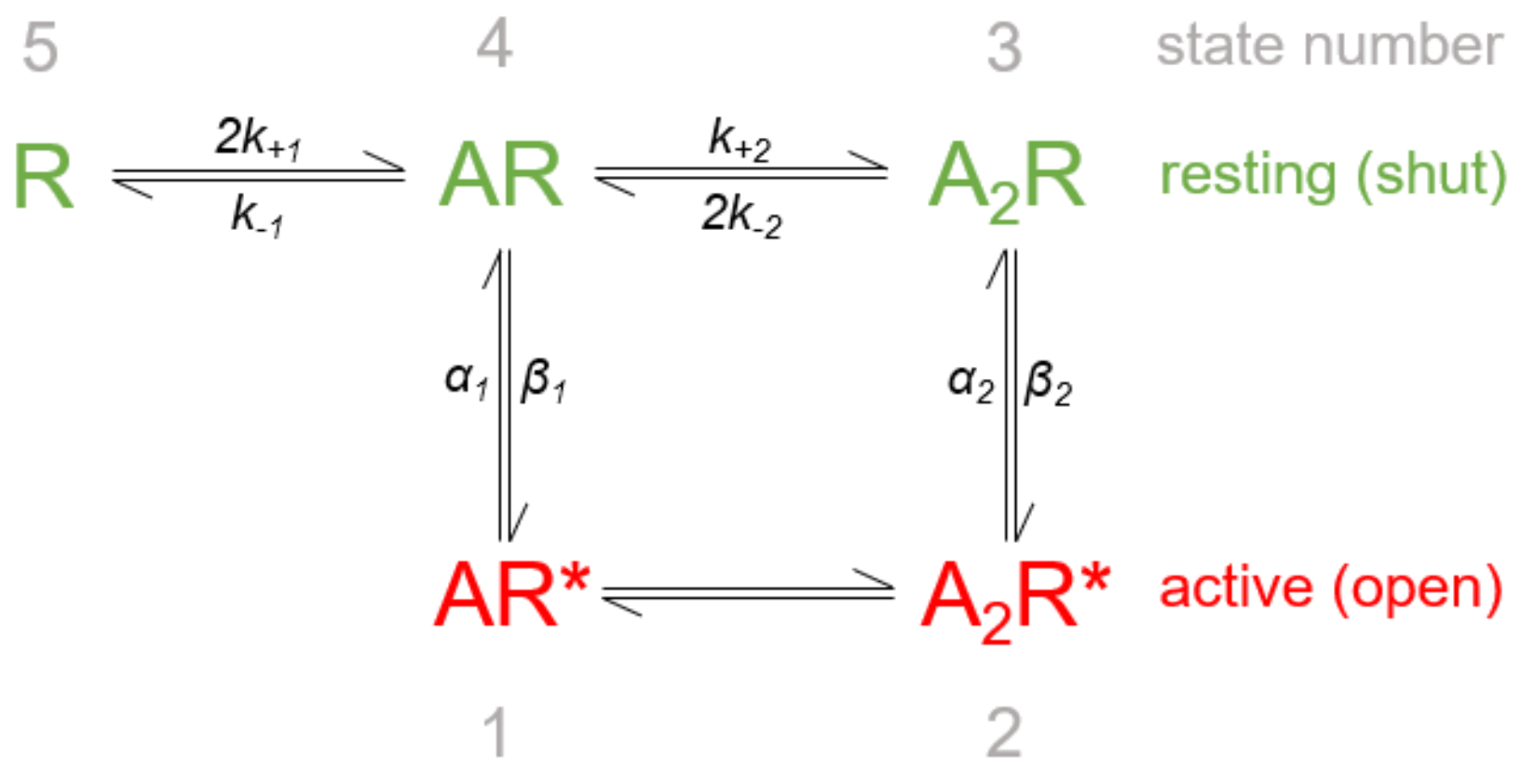

(56)

This model is useful for checking the calculations because it was used in [6] and [23] as a numerical example. The numerical results in the references can be checked against the values produced by the R script, `CK-kinetics.R`, provided in the electronic supplementary material. Use model 3, CH82, with rate set "`CH82 (orig rates)`", to reproduce the numbers in [6] and [23] . Those examples were calculated with a low agonist concentration, 100 nM, because they were intended to illustrate single ion channel experiments. For the purposes of illustrating a ligand binding experiment, we use a higher concentration, and the rate constants were defined such that both binding sites have the same affinity for the agonist. The results in Figure 7 were calculated using `CK-kinetics.R` with model 3 (CH82) and the rate set "`equal bindings`". The results are in the printout file, "`CH82 with equal bindings v5.7.txt`", in the electronic supplementary material.

The graphs in Figure 7 were calculated using the following values for rate constants. The first and second ligand binding events are the same: both have dissociation rate constant $k_{-} = 2000\ \mathrm{s}^{-1}$ and association rate constant $k_{+} =$ 5 x $10^8\ \mathrm{M}^{-1}\mathrm{s}^{-1}$. The equilibrium constant for both binding sites is the ratio of these, $K_\mathrm{A} = 4\ \mu M$. Opening and shutting rate constants for the diliganded receptor are $\beta_2 = 15000\ \mathrm{s}^{-1}$ and $\alpha_2 = 1000\ \mathrm{s}^{-1}$, so the efficacy, $E_2 = \beta_2/\alpha_2 = 15$, is quite high (maximum response = 15/(1 + 15) = 94%). The efficacy for monoliganded receptors is much lower, so monoliganded openings are rare.

Figure 7A shows the response to a step change in concentration from 0 to $5\ \mu\mathrm{M}$ at $t = 0$ followed by a step back to zero concentration at $t = 5\ \mathrm{ms}$. There is an initial fast rise in monoliganded state (AR, purple line), and the second binding event follows ($A_2R$, green line), but the fast opening rate of diliganded receptors means that, once in $A_2R$, opening is much more likely than dissociation and the diliganded open state, $A_2R^*$ (red line) soon predominates.

The time course of ligand binding is shown in Figure 7B. It's calculated as the sum of $k - 1 = 4$ exponential components, with the same time constants as the curves in Figure 7A (see the printout "`CH82 with equal bindings v5.7`").

$$p_{\text{bound}}(t) = 0.919 + 0.0048 \exp\left(-\frac{t}{0.049}\right) - 0.192 \exp\left(-\frac{t}{0.129}\right)$$
$$-0.003 \exp\left(-\frac{t}{0.182}\right) - 0.729 \exp\left(-\frac{t}{0.666}\right) \tag{57}$$

The onset of ligand binding is dominated by the slowest component, with time constant 0.666 ms, though ligand binding is initially faster than this (Figure 7B) because the amplitude of the second exponential term, with time constant 0.129 ms, is large enough to make a difference, though it would need precise measurements to detect it. Although the amplitude of the second slowest component is more than 25% of the amplitude of the slowest ($0.192/0.729 = 0.263$), it is still not big enough to be easily detected and fitted. This illustrates the limitations of fitting multiple exponentials to experimental observations.

The offset of ligand binding is almost entirely described by the slowest component, with a time constant of 4.65 ms (see Figure 7B and the accompanying printout).

Insofar as the onset and offset of ligand binding are fitted with single exponentials, Eq. (9) can be used to provide an estimate of the apparent equilibrium constant for binding from the measured rates. In this case we find $K_{\text{app}} = 0.835\ \mu\text{M}$. This is a long way from the microscopic equilibrium constant for binding, $4\ \mu\text{M}$, but it is not greatly different from the $EC_{50}$ for the equilibrium binding curve, $1\ \mu\text{M}$. The concentration of agonist required to occupy 50% of the binding sites for the mechanism in Scheme (56) is

$$EC_{50} = \sqrt{\frac{K_{\text{A1}}K_{\text{A2}}}{1 + E_2}} \tag{58}$$

where $K_{\text{A1}}$ and $K_{\text{A2}}$ are the equilibrium constants for the two distinct binding sites. In the present example, these are the same (both are 4 μM), so Eq. (58) reduces to

$$EC_{50} = \frac{K_{\text{A}}}{\sqrt{1 + E_2}} \tag{59}$$

The analogy between this result and the effective equilibrium constant for the dCK mechanism, Eq. (31), is obvious. This could equally be denoted as the effective, or macroscopic, equilibrium constant for ligand binding, $K_{\text{eff}}$, for the CH82 model. In the present example, this gives $EC_{50} = 1\ \mu\text{M}$. These numbers are given at the end of the printout file, "`CH82 with equal bindings v5.7`". It would take further investigation to define the range of conditions under which $K_{\mathrm{app}}$, calculated from the onset and offset rates using Eq. (9), is close to the $EC_{50}$ for ligand binding.

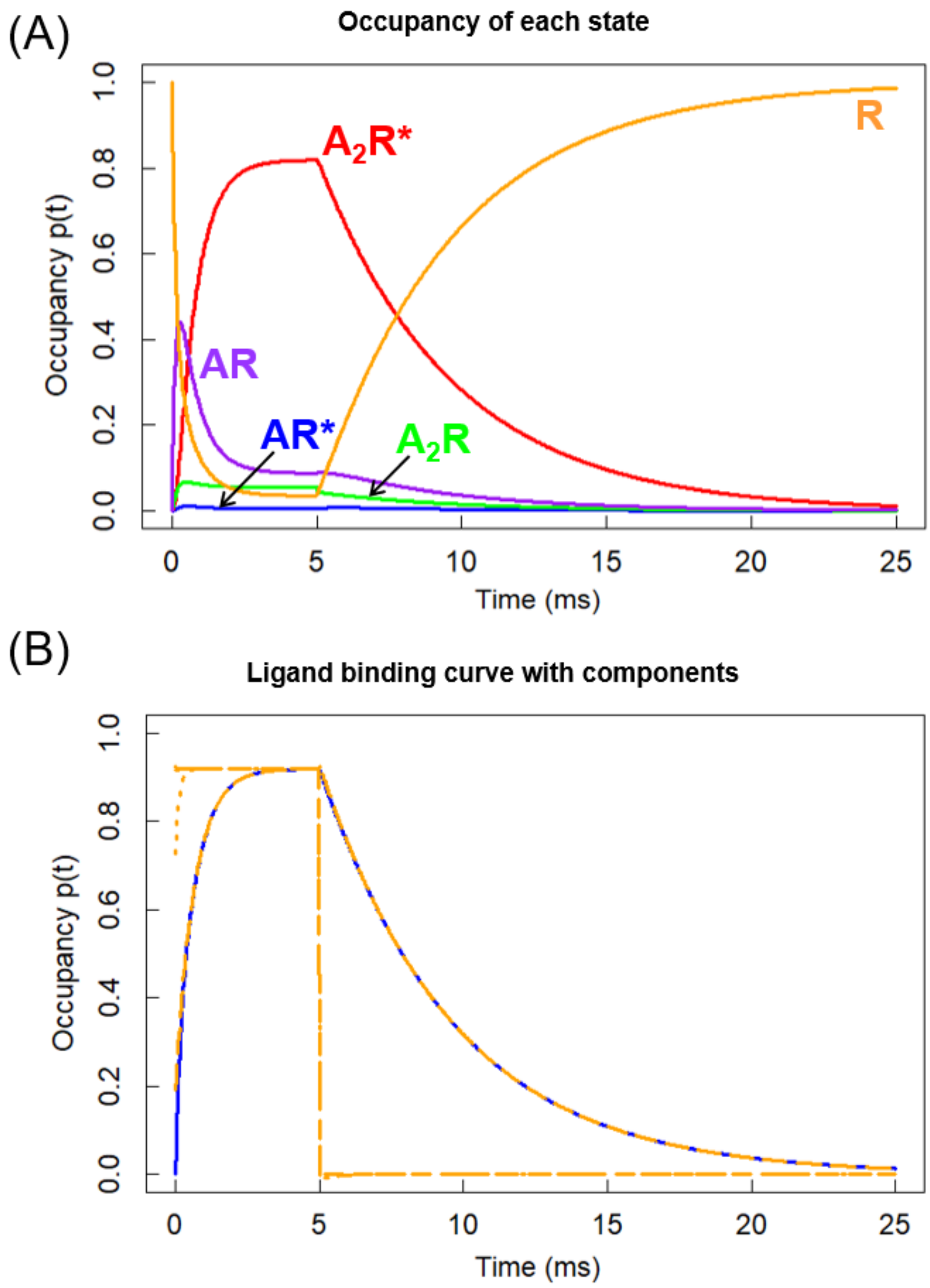

**Figure 7.** An example of the CH82 model for the muscle-type nicotinic receptor shown in Scheme (56). (A) The time course of the occupancy of each of the states in the CH82 mechanism during a 5 ms pulse of agonist (5 μM). (B) The time course of ligand binding during a 5 ms pulse of agonist. The observed time course of binding is shown in blue, with the components shown in orange. Plots are produced by `CK-kinetics.R`. For details, see the printout file, "`CH82 with equal bindings v5.7`" in the electronic supplementary material.

# 5. Discussion

## 5.1. The simple case

The problem of understanding the kinetics of ligand-receptor interactions was first dealt with by AV Hill [7] who derived expressions describing the binding of curare to binding sites on skeletal muscle. Hill showed that, for a simple binding reaction, as in Scheme (1), the binding of a ligand will approach equilibrium following a single exponential time course. He showed that observation of the macroscopic rate of ligand binding at different ligand concentrations allows the estimation of $k_{+1}$ and $k_{-1}$, and, hence, the microscopic affinity.

It's worth noting that attempts to measure the rates of binding and dissociation of ligands have often been frustrated by the limitations imposed by diffusion through tissues. AV Hill's own measurements suffered from this problem. To make matters still harder, it's possible for the ratio of rate constants to agree roughly with an independently measured equilibrium constant, even when the rate constants are both too slow. This can happen, roughly, when access to the binding site is slowed by diffusion, and diffusion is slowed by binding of the ligand [35].

Hill's approach is valid only for ligands which bind to their receptor and do not evoke any change in the receptor's conformation. If the binding of the ligand causes the receptor to change conformation, as is the case for an agonist, then the reaction can no longer be described by a simple binding reaction. Yet, Hill's approach is still often used to analyse the kinetics of agonist binding. In the simplest case, that of the dCK mechanism, the kinetics of agonist binding will be described by the sum of two exponential components with macroscopic rates given in Eq. (25). So, even in this simple case, the binding of an agonist is too complicated to allow for such simple estimation $k_{+1}$ and $k_{-1}$ because binding depends on *all* of the microscopic rate constants in the reaction mechanism.

## 5.2. Fitting exponentials to data

There are several cases in which the rate of ligand binding has been observed to need more than one exponential component to describe it. But we can find no examples of cases in which microscopic rate constants have been estimated from observations by fitting experimental results. An obvious procedure would be to do a least-squares fit, weighted if necessary, to the observed curves with the free parameters being the microscopic rate constants in a specified mechanism, rather than the time constants and amplitudes of the observed exponential components. One advantage of this procedure is that the mechanism would dictate the number of components that were fitted, thus circumventing the need to use (mostly unsatisfactory) methods for deciding how many exponential components to fit.

It's important that a fitting procedure provides estimates of the uncertainty in the estimated rate constants, as a precaution against over-fitting. When the data consist of discreet, independent data points, the methods for estimating an approximate covariance matrix and, for example, likelihood intervals, are standard: see Beale [36],

Efron & Hinkley [37] and Colquhoun & Sigworth [38]. But when the data consist of 'continuous' curves, *i.e.,* data points sampled at a high rate, the points will be serially correlated and that will result in the effective number of data points being overestimated and, consequently, errors will be underestimated. Methods of dealing with this problem are described in [39].

### 5.3. An instructive example

An instructive example is that of the binding of nitric oxide (NO) to soluble guanylate cycle (sGC), investigated by Zhao *et al.* [40,41] using a stopped-flow method with spectroscopic measurements. In their experiments, Zhao and colleagues identified two distinct exponential phases in the binding of NO to sGC. The authors assigned these two exponential phases to the binding of NO (the fast component) and the subsequent activation of sGC (the slow component). But, as we have seen, when there is more than one component in the time course of ligand binding, those components do not, in general, correspond to any specific step in the reaction. Both components depend on *all* of the microscopic rates in the mechanism (*e.g*. see Eq. (23)-(25)). There is no one-to-one relationship between the observed rate of ligand binding and the microscopic rate constants describing given steps in a reaction mechanism.

The authors suggested a novel reaction mechanism based on their finding that the second, slower, exponential phase was dependent on the concentration of NO. However, Bellamy *et al.* [42] pointed out that there is no need for such complexity because the simple dCK mechanism predicts that both exponential components will be dependent on the concentration of NO. None of these papers used fitting procedures like those described above. Such optimum analyses seem to have been used only for analysis of single ion channel recordings [43,44], where they have been tested by simulations [12].

### 5.4. Why only one exponential component may be detected

We return now to the question of what is estimated when models with more than two states are treated as though they had only two. In particular, what is the result of using Eq. (9) to estimate the equilibrium constant for binding? This can be done only if the onset and offset of binding are described by a single exponential, so that $k_{\mathrm{on}}$ and $k_{\mathrm{off}}$ in Eq. (9) are defined unambiguously by the data. This is illustrated most simply by the dCK model. This predicts that the time course of receptor occupancy will be described by the sum of two exponential components. And, if two components are visible, as in the example in Section 3.3.1, it isn't possible to use Eq. (9).

Quite often, though, the relaxations to equilibrium are quite close to being a single exponential. For example, synaptic currents are often close to showing a single exponential decay, and, insofar as the transmitter concentration decays rapidly, the synaptic current can be treated as a relaxation after a step to zero concentration. Bear in mind that the time course of the number of open ion channels has the same time constants as the time course of ligand binding – only the amplitudes differ. It was postulated in 1977 that the time constant for the decay of muscle endplate currents was given by the mean length of the short burst of openings elicited by the very brief pulse of transmitter [5]. Insofar as the length of such bursts is close to being exponentially-distributed, the decay of the synaptic current will be a simple exponential, see [25] (especially Section 9.3.). The relationship between single channel burst length and the macroscopic relaxation after a jump to zero agonist concentration was proved formally by Wyllie and colleagues [34].

### 5.5. Cases of the dCK mechanism in which a single exponential predominates

The three special cases of the dCK mechanism in which both onset and offset are close are close to being described by a single exponential are discussed in Section 3.2, with numerical examples of each case in Section 3.3.

The case in which binding is fast relative to the speed of the subsequent conformation change (Section 3.2.1, Figure 1A), was exemplified in Section 3.3.2 and Figure 3. The measurable time constants for onset and offset, when the corresponding values of $k_{\text{on}}$ and $k_{\text{off}}$ are substituted into Eq. (9), are expected, according to Eq. (36), to give a concentration-dependent value with no direct relationship to an equilibrium binding constant. In the numerical example, Eq. (36) gives quite a good approximation, but the value isn't interpretable as a binding affinity. In the case of fastish binding, exemplified in Section 3.3.3 and Figure 4, two exponential components are visible, so Eq. (9) can't be used.

The case in which the conformation change is fast was discussed in Section 3.2.2 and Figure 1B, and exemplified in Section 3.3.4 and Figure 5. In this case, it's expected that the values of $k_{\text{on}}$ and $k_{\text{off}}$, when substituted in Eq. (9), give an estimate of $K_{\text{eff}}$, as shown in Eq. (42). This is the equilibrium constant for binding that would be measured in an equilibrium binding experiment, as the concentration of ligand that results in half the binding sites being occupied at equilibrium, the $EC_{50}$. Since this depends on the microscopic equilibrium constants for both binding and for activation, it's impossible to separate affinity and efficacy from measurements at equilibrium.

The third case in which a single exponential component dominates both the onset and offset is the case in which the concentration of the intermediate state, AR, is low. This was discussed in Section 3.2.3 and exemplified in Section 3.3.5 and Figure 6. In this case too, the rates of onset and offset are expected to estimate $K_{\text{eff}} = K_{\text{A}}/(1+E)$ only when the efficacy, $E$, is much larger than one, see Eq. (48). Since this case assumes that the activation rate, $\beta$, is fast, this implies that the efficacy, $E = \beta/\alpha$, is large.

The mechanism discussed in Section 4 is more complicated than the dCK model. It involves the binding of two ligand molecules in order to fully activate the receptor. With the numbers chosen for the numerical example, that mechanism also exhibits time courses for the onset and offset of ligand binding that are predicted to be close to single exponentials (see Figure 7). In this case too, the onset and offset rates for ligand binding, when used in Eq. (9), give a value quite close to the $EC_{50}$ for ligand binding, which is given in Eq. (59). We haven't tested a wide enough range of rate constants to see how generally this result would hold. That would be straightforward to do with the R script provided in electronic supplementary material.

## 6. Conclusions

When the rates of ligand binding are measured by methods like surface plasmon resonance, it's common practice to use the observed rate constants for onset and offset to estimate an equilibrium constant for ligand binding. If this agrees with the equilibrium constant found as the $EC_{50}$ for equilibrium binding, this is taken as validation of the measured rates. The method for inferring an equilibrium constant from rates assumes a simple binding reaction that produces no conformation change. This can never be so for agonists which, by definition, produce a conformation change in the receptor to which they bind.

Our investigation of two models that include a conformation change after binding lead to the following conclusions:

1. All such models predict that relaxation towards equilibrium will be described by more than one exponential component.
2. Nonetheless, it is not uncommon that the onset and offset of binding are dominated by one exponential component, *e.g.* one that has a larger amplitude than all of the other components.
3. In such cases, the rates of onset and offset may be fitted tolerably well by single rate constants. By treating the system as a simple binding reaction, these two rates can be used to estimate an equilibrium constant, as in Eq. (9).
4. This equilibrium constant is never the microscopic equilibrium constant, $K_{\mathrm{A}}$, for the binding step of the reaction mechanism, which provides information about the agonist binding site.
5. In the case of the simple dCK mechanism, the estimated equilibrium constant may be close to the macroscopic equilibrium constant for ligand binding, $K_{\mathrm{eff}} = K_{\mathrm{A}}/(1+E)$. This is the concentration of ligand needed to occupy half of the binding sites at equilibrium. It depends on both of the underlying microscopic equilibrium constants in the mechanism, so it doesn't allow the separate measurement of affinity ($K_{\mathrm{A}}$) and efficacy ($E$).
6. When binding is much faster than the subsequent conformation change, there will be only one measurable rate constant for offset and onset, but the rates of onset and offset would not give an estimate of $K_{\mathrm{eff}}$, so agreement with the measured $EC_{50}$ would not be expected. In this case, if the amplitude of the initial fast jump in binding could be measured, it would provide an estimate of the microscopic binding constant, $K_{\mathrm{A}}$.
7. When the conformation change is much faster than binding, it is expected that measurements of onset and offset rates would yield, via Eq. (9), an estimate of $K_{\mathrm{eff}} = K_{\mathrm{A}}/(1+E)$. So, in this case, agreement between this value and the observed $EC_{50}$ would be expected.

8. In the third case in which the dCK mechanism would give rise to approximately single exponential time courses for the onset and offset of ligand binding, the case of low intermediate concentration, it is expected that the equilibrium constant estimated from onset and offset rates would be $K_A/E$, which would agree with the $EC_{50}$ only for high efficacy agonists, for which $E$ is large.
9. If the time course of the onset and offset of ligand binding is approximately described by a single exponential component, it is, in principle, possible to learn more from the concentration-dependence of onset rates, but for this to be feasible, you'd need to postulate a realistic mechanism.
10. If more than one exponential component can be measured, it is, in principle, possible to estimate the microscopic rate constants in the underlying mechanism, but we are unaware of any attempt to do this from ligand binding measurements.

**Electronic supplementary material**

The supplement to this paper, including the R script and details of each of the examples in Section 3.3, is available at:

https://doi.org/10.6084/m9.figshare.33406285